# Image Retrieval And Classification Using Local Feature Vectors

*A Project Report*

*submitted by*

**VIKAS VERMA**

*in partial fulfillment of the requirements*
*for the award of the degree of*

**MASTER OF TECHNOLOGY**

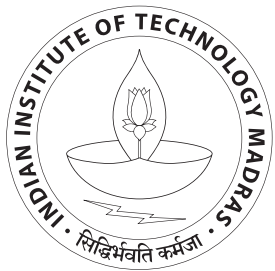

**DEPARTMENT OF COMPUTER SCIENCE & ENGINEERING**
**INDIAN INSTITUTE OF TECHNOLOGY MADRAS.**
**June 2011**

# THESIS CERTIFICATE

This is to certify that the thesis entitled **"Image Retrieval And Classification Using Local Feature Vectors"**, submitted by **Vikas Verma**, to the Indian Institute of Technology, Madras, for the award of the degree of **Master of Technology**, is a bonafide record of the research work carried out by him under my supervision. The contents of this thesis, in full or in parts, have not been submitted to any other Institute or University for the award of any degree or diploma.

June 2011

(Dr. C. Chandra Sekhar)
Department of Computer Science & Engineering
Indian Institute of Technology Madras
Chennai, Tamilnadu 600036

# ACKNOWLEDGEMENTS

First and foremost, I express my sincere gratitude to Dr. C. Chandra Sekhar for his guidance throughout my research work. His invaluable advise and constructive criticism, patience and constant encouragement have motivated me throughout my research work. Although he was very busy with the administration work, he was always available to steer my project in methodical way. His clear thought process and lucid communication made me understand the problem easily. I also thank him for advice, comments and keen observations during the Monday and Wednesday Laboratory meetings.

I take this opportunity to thank the faculty advisors, Dr. Janacek, Dr. C. Pandu Rangan and Dr. Madhu Mutyam for ensuring smooth sailing over the entire duration of my stay in IIT. I thank the head of department Dr. C. Siva Ram Murthy for providing excellent computing and library facilities in the department for project work.

I also express my gratitude towards my lab members Dileep A.D., Veena T., Yogesh Bendre and Shaji Mohan for sharing their experience with me. My special thanks to Bala Sanjeevi for always being there to help selflessly.

Finally, I would like to express my deepest sense of gratitude to my family and friends for their support and understanding.

# ABSTRACT

Content Based Image Retrieval(CBIR) is one of the important subfield of Information Retrieval field. The goal of a CBIR algorithm is to retrieve semantically similar images in response to a query image submitted by the end user. CBIR is a hard problem because of the phenomenon known as *semantic gap*.

In this thesis, we aim at analyzing the performance of a CBIR system built using local feature vectors and Intermediate Matching Kernel. We also focus on reducing the response time and improving the retrieval performance of a CBIR system. We have proposed a Two Step Matching process for reducing the response time and developed a Meta-Learning framework for improving the retrieval performance. Results have shown that Two Step Matching process significantly reduces response time and Meta-Learning Framework is able to improve the retrieval performance by more than two times. To this end, we have analyzed the performance of different image classification systems which use different image representation constructed from the local feature vectors.

# Contents

# List of Tables

# List of Figures

# List of Algorithms

# ABBREVIATIONS

| | |
|---|---|
| **AP** | Affinity Propagation |
| **BOW** | Bag-of-words |
| **CBIR** | Content Based Image Retrieval |
| **GMM** | Gaussian Mixture Model |
| **IMK** | Intermediate Matching Kernel |
| **kNN** | k-Nearest Neighbors |
| **LDA** | Latent Dirichlet Allocation |
| **MAP** | Mean Average Precision |
| **PDF** | Probability Density Function |
| **SVM** | Support Vector Machine |

# Chapter 1

# CONTENT BASED IMAGE RETRIEVAL- AN INTRODUCTION

## 1.1 Definition and Motivation

Any technique that helps to systematize or organize digital images by their visual content can be regarded as CBIR system. That is, any technique ranging from an simple image similarity function to robust image search engine (like Google image search) falls under the category of CBIR. The goal of CBIR systems is to operate on image data and, in response to a visual a query, extract relevant images from the dataset. This process is shown in Figure 1.1. Note that feature extraction from query image and similarity measurement are online processes while feature extraction from database images is offline process.

The recent tremendous growth in internet and digital technologies have brought a substantial increase in amount of available digital imagery. Storage of such image data is relatively straight-forward, but easy searching and retrieval of such data imposes the need to have a system that can efficiently and effectively organize this data. This need to have a versatile and general purpose system for organizing large image database is the motivation behind Content based image retrieval (CBIR) system.

CBIR techniques consist of diversified areas like image segmentation, feature extraction, feature representation, storage and indexing, image similarity measurement and retrieval. All these things make developing a CBIR system, a challenging task.

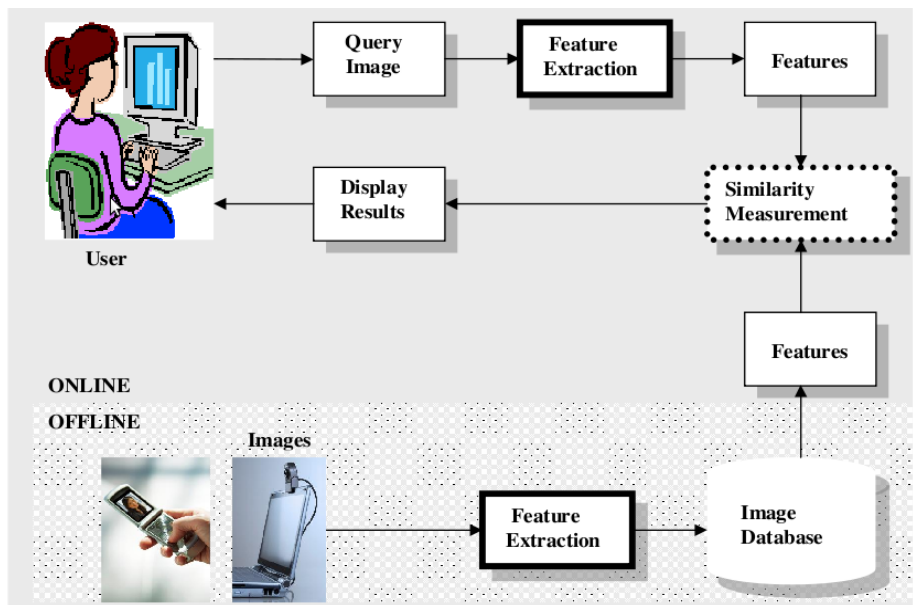

Figure 1.1: Processes involved in CBIR

## 1.2 Challenges Involved

### 1.2.1 Semantic Gap

Semantic of an image refer to meaning of the image content. In contrast to low-level image visual features, semantic is high level concept. Often images having similar low-level features may contain different concept, e.g.,an image of The Sun and an image of an orange both have similar color features, but they contain entirely different concept. Similarly images having different low-level visual features may contain same concept, e.g., images of objects like vehicle may have different color and shape, hence different low-level feature, still they represent same concept. This lack of coincidence between the low-level features and high level concepts present in the images is known as *Semantic gap*. One of the most important challenges of CBIR systems is how to bridge this semantic gap.

### 1.2.2 Computational Cost

In context of CBIR systems, we define computational cost as the elapsed time between the submission of query by user and the retrieval of results by the CBIR system. Along with satisfactory retrieval results, the system should provide results to end user in reasonable amount of time. Several methods have been proposed in literature to decrease the computation cost of image retrieval. Broadly, these methods fall under two categories as follows:

- First category consist of methods which aim to reduce the dimensionality of feature vector which in turn results in lesser computation cost during distance calculation for similarity retrieval.

- Second category consist of methods which aim to select most important feature or combination of features among available features.

### 1.2.3 Usability

Usability of CBIR system is defined as the ease with which end user is able to use the system. It is also associated with the scalability and adaptability of the system. Scalability refers to system's capability of handling changing amount of media and query load. Adaptability refers to the system's ability to adapt according to changes in environment. A system is said to be more adaptable if its performance is not affected by change in image database, platforms and hardware architecture.

## 1.3 Objective and Scope of this Work

A large number of approaches for CBIR and image classification have been proposed to date. They can be broadly categorized in two types: one that use global feature vector and another that use local feature vectors, for computing similarity between a pair of images. The objective of this work is to build a CBIR system and an image classification system using local feature vectors (or varying length patterns). Through this work we have made following contribution:

- We have developed One Step Matching and Two Step Matching based CBIR system using Intermediate Matching Kernel.

- In Two Step Matching approach we have analyzed the effect of using different clustering algorithms during first step. We have also proposed to consider more than one nearest cluster as potential search space in Two Step Matching approach.

- We have developed a Meta-learning framework for CBIR that is shown to have considerably improve the performance of underlying CBIR system.

- We have build bag-of-words based, Latent Dirichlet Allocation based and IMK based image classification systems and compared their performances.

## 1.4 Organization of the Thesis

The thesis is organized as follows :

- **Chapter 2** describes background and related work of the problem.

- **Chapter 3** describes use of Intermediate Matching Kernel in One Step Matching and Two Step Matching approaches of Content based image retrieval (CBIR).

- **Chapter 4** describes a Meta-learning framework for improving the performance of any existing CBIR system.

- **Chapter 5** describes different techniques for classification of images using varying length patterns.

- **Chapter 6** describes summary and conclusion of this work.

# Chapter 2

# BACK GROUND AND RELATED WORK

The field of CBIR has grown tremendously in last two decades. It has also found applications in practical life as it is used in several commercial, governmental, and academic institutes such as libraries, TV broadcasting channels, governmental archives. Content-based image searching, browsing and retrieval applications are required for users from various domains such as remote sensing and surveillance. CBIR is primarily based on the visual features of images such as color, texture and shape information. Numerous methods have been developed to date to extract image content characteristic from the visual data automatically, and to use the extracted image content information for retrieval of images in response to query image.

Rest of the chapter is organized as follows: In Section 2.1 we present the survey of field of CBIR. Section 2.2 explains various learning techniques used for accomplishing the task of image retrieval. Image features used for image retrieval are explained in Section 2.3. Section 2.4 explains the method for creating image representation using local feature vectors. Kernel methods for the task of image retrieval and classification are explained in Section 2.5. Finally, Section 2.6 and 2.7 present the performance measure for the task of image retrieval and image classification.

## 2.1 Literature Survey

Numerous methods have been proposed for CBIR. A semantics-sensitive approach to content-based image retrieval has been proposed in SIMPLIcity (5). The matching measure, termed integrated region matching (IRM), has been constructed for faster retrieval using region feature clustering and the most similar highest priority (MSHP) principle(6). Region based image retrieval has also been extended to incorporate spatial similarity using the Hausdorff distance on finite sized point sets (7), and to employ fuzziness to characterize segmented regions for the purpose of feature matching(8). A framework for region-based image retrieval using region codebooks and learned region weights has been proposed in (9). A new representation for object retrieval in cluttered images without relying on accurate segmentation has been proposed in (10). Another perspective in image retrieval has been region-based querying using homogeneous color-texture segments called blobs, instead of image to image matching(41).

Instead of using image segmentation, one approach to retrieval has been the use of hierarchical perceptual grouping of primitive image features and their inter-relationships to characterize structure(12). Another proposition has been the use of vector quantization (VQ) on image blocks to generate codebooks for representation and retrieval, taking inspiration from data compression and text-based strategies (13). A windowed search over location and scale has been shown more effective in object-based image retrieval than methods based on inaccurate segmentation (14). A hybrid approach involves the use of rectangular blocks for coarse foreground/background segmentation on the users query region-of-interest (ROI), followed by the database search using only the foreground regions (15).

## 2.2  Different Learning Techniques and Their Application to Image Retrieval

It was not late when researchers start realizing that it is too ardent to use a single similarity measure to develop a system that produces robust semantic based ranking of images. Thus attempts have been used to harness the ability of learning based techniques for the task of image retrieval. In this section we present these techniques, their use and limitations.

## 2.2.1 Image Classification

In context of image retrieval, image classification has often been used as a pre-processing step for reducing the response time to query image in large databases and improving accuracy. More elaborately, in image retrieval systems, an query image is classified into one of category in database predicted class, subsequently, a similarity measurement step is carried out over only those images that belong to the same visual category as predicted for the query image.

Image classification is applicable only when labeled training images are available. Domain-specific database such as medical images database, remotely sensed imagery are example of databases where labeled training images are readily available. Classification methods can be partitioned into two major branches: discriminative classification approach, generative classification approach. In discriminative approach decision boundaries are estimated directly, e.g., SVM and dicision trees. This approach does not need any prior information about classes. In generative approach, the density of data with in each class is estimated separately and Bayes formula is then used to compute the posterior on test data. This approach is easier to incorporate any prior information and more efficient when there are many classes.

For the purpose of image retrieval Bayesian classification is used in Vailaya et al (19). Classification based on SVMs has been applied to images in Goh et al (20). SVM based supervised learning can be also used for preprocessing of multimedia queries as shown in Panda and Chang (21) Image classification based on a generative model is explored in Datta et al (22).

## 2.2.2 Image Clustering

When labeled data is not available, unsupervised clustering can be useful for speeding up the retrieval process. Image clustering is specifically applicable to web image data where meta data is also available for exploitation in addition to visual features (23; 24; 25).

Clustering techniques can be partitioned into three categories: pair-wise distance based, optimization of an overall clustering quality measure and statistical modeling. The pair-wise distance based methods, e.g., linkage clustering and spectral graph partitioning do not depend on the mathematical representation of data instance, hence they have general applicability. They are particularly useful in image retrieval because image representation may be often very complex. However, they have one disadvantage of having high computational cost because we need to compute an order of $n^2$ pair-wise distances, where $n$ is the size of the data. In CBIR systems, similarity information among the retrieved images is also available for exploitation. In this respect, Chen et al (26) proposes the use of a new spectral clustering-Shi and Malik (27) based approach to incorporate such information into the retrieval process.

Clustering techniques based on the optimization of an overall measure of the clustering quality is a common approach explored since the early days of pattern recognition. The simple and highly popular method, k-means clustering, is one example of this type of clustering. In k-means, the quality of clustering result is measured by the sum of within-cluster distances between every data instance and its cluster centroid. Here, if the number of clusters is not specified, a simple method to determine this number is to gradually increase it until the average distance between a vector and its cluster centroid is below a given threshold. In Gordon et al.(28), an unsupervised clustering approach for images has been proposed using the Information Bottleneck (IB) principle. The proposed method works for discrete (histograms) as well as continuous (Gaussian mixture) image representations.

In statistical modeling based clustering techniques, the general idea is to treat every cluster as a pattern characterized by a distribution, and the overall data set is thus a mixture of these distributions. For continuous vector data, the most used distribution of individual vectors is the Gaussian distribution. By fitting a mixture of Gaussians to a data set, usually by the EM algorithm McLachlan and Peel (29), we estimate the means and covariance matrices of the Gaussian components, which correspond to the center locations and shape of clusters.

### 2.2.3 Relevance Feedback

The idea behind relevance feedback(RF) is to take the results that are initially returned from a given query and to use information about whether or not those results are relevant to perform a new query. Essentially, RF is a query modification technique which attempts to capture the users precise needs through iterative feedback and query refinement. It can be thought of as an alternative search paradigm, complementing other paradigms such as keyword based search or example based search. In the absence of a reliable framework for modeling high-level image semantics and subjectivity of perception, the users feedback provides a way to learn case-specific query semantics.

RF provides a trade-off between a fully automated, unsupervised system and one based on the abstract user needs. Although query refinement is an attractive proposition when it comes to a very diverse user base, there is also the question of how well the feedbacks can be utilized for refinement. Whereas a user would prefer shorter feedback sessions, there is an issue as to how much feedback is enough for the system to learn the user needs.

## 2.3 Image Features

In literature, a number of image features have been proposed ,e.g., color histogram, Tamura Features, Global Texture Descriptor, Gabor histogram, Gabor vector, Invariant Feature Histograms, MPEG 7: scalable color, MPEG 7: color layout, MPEG 7: edge histogram. Description and experimental comparison of these features is presented in Deselaers et al (16). Moreover, an image can be represented using global feature extracted from whole image or local features extracted from parts of image.

### 2.3.1 Motivation Behind Using Local Feature Vectors of Images

The global feature is sensitive to change due to perceptive distortion, occlusions, illumination variations and clutter. It may also be redundant, contain misleading information and fail to capture finer semantic details of an image (2). Local features are powerful image descriptors and provide better image representation when compared to global features (3).

### 2.3.2 Extraction of Local Features

Local features can be extracted from an image using following three approaches:

- Image is segmented into regions and from each region a feature vector is extracted.

- Salient point in images are detected and a feature vector is extracted from a region around each of these salient points.

- Image is partitioned into fixed size blocks and from each block a feature vector is extracted.

These approaches are shown in Figure 2.1:

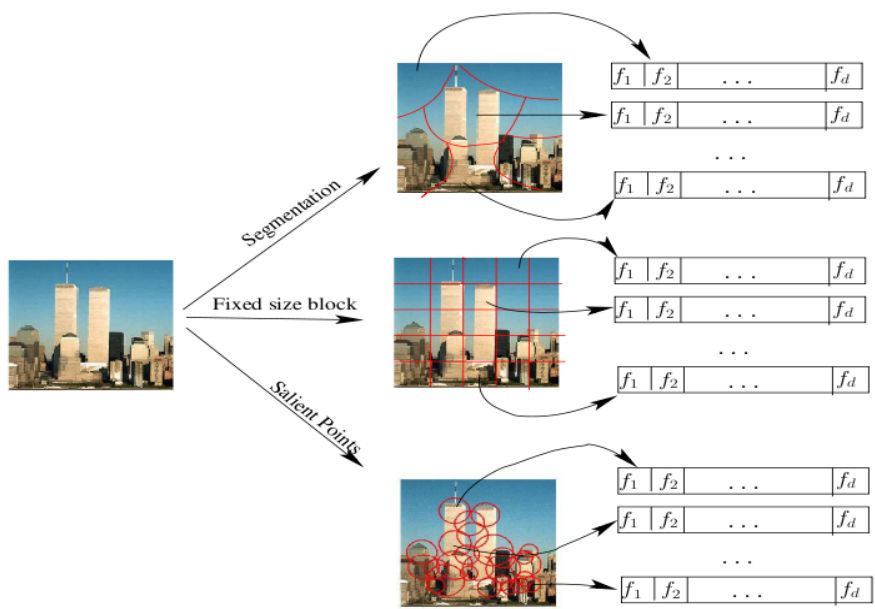

Figure 2.1: Different approaches for extracting local features

## 2.4 Image Representation Using Local Feature Vectors

Once the local feature vectors are extracted from an image, the next step is to create a representation of the image using these feature vectors. This representation is also called *image signature*. Let the local features extracted from an image be denoted as $\mathbf{x}_i \in \mathbb{R}^d, i = 1, ........n$, where $d$ is dimensionality of feature vector. The various ways of creating image representation using local feature vectors are described below:

### 2.4.1 Super-vector of Local Feature Vectors

When the number of local feature vectors extracted from different images are same, the easiest way to create a image representation is to concatenate all the local feature vectors to form a super-vector as

$$\mathbf{X}_{super} = [\mathbf{x}_1, ........, \mathbf{x}_n] \qquad (2.1)$$

The dimension of $\mathbf{X}_{super}$ is $d \times n$. Super-vector representation of image inherently contains some spatial information, since local feature vectors are concatenated in some fixed order.

### 2.4.2 Probabilistic Modeling of Local feature Vectors

In this method, local feature vectors from each image are modeled using a probability distribution. Gaussian Mixture Model (GMM) are the most commonly used model for this purpose (41; 42; 43). For proper estimation of parameters of GMM it is necessary to have enough number of local feature vectors.

### 2.4.3 Bag-of-Visual Words Representation

In this representation, an image is represented as a vector of frequency of occurrences of visual words in image.For construction of visual-words, local feature vectors of all images are clustered into large number of clusters and centers of these clusters are considered to be visual words. Although, this method maps local feature vectors of an image onto a fixed dimensional representation, it has a limitation that significant amount of information may loss because this method involves vector quantization of local feature vectors. More detailed explanation of this method is given in Section 5.2.

### 2.4.4 Set of Feature Vectors Representation

All the above representations have some limitations. A more robust way to represent image using local feature vector is to represent it as set of local feature vectors, given by:

$$\mathbf{X}_{set} = \{\mathbf{x}_1, ........., \mathbf{x}_n\} \qquad (2.2)$$

This representation has advantage that unlike bag-of-word representation ,it does not incur any information lose. Also, unlike probabilistic model based representation it does not require any assumption about the distribution of local feature vector.

When images are represented as set of local feature vectors, it calls for a method that can compute similarity/dissimilarity between a pair of set of local feature vectors. In next section we discuss such methods.

## 2.5 Kernel Methods for Image Retrieval and Classification using Set of Local Feature Vectors

Let $\mathbf{X}_i = \{\mathbf{x}_{i1}, \mathbf{x}_{i2}, ..., \mathbf{x}_{iM_i}\}$ and $\mathbf{X}_j = \{\mathbf{x}_{j1}, \mathbf{x}_{j2}, ..., \mathbf{x}_{jM_j}\}$ be the sets of local feature vectors for the examples $\mathbf{X}_i$ and $\mathbf{X}_j$ respectively.

### 2.5.1 Summation Kernel

The summation kernel (38) is computed by matching every local feature vector in $\mathbf{X}_i$ with every local feature vector in $\mathbf{X}_j$ as follows:

$$K^S(\mathbf{X}_i, \mathbf{X}_j) = \sum_{m=1}^{M_i} \sum_{n=1}^{M_j} k(\mathbf{x}_{im}, \mathbf{x}_{jn}) \qquad (2.3)$$

where $k(.,.)$ is a basic Mercer kernel for two local feature vectors that gives a measure of similarity between them. The summation kernel is a Mercer kernel. The number of basic kernel computations is $M_i * M_j$ for this kernel.

## 2.5.2 Matching Kernel

The matching kernel (39) is constructed by considering the closest local feature vector of an example for each local feature vector in the other example as follows:

$$K^{MK}(\mathbf{X}_i, \mathbf{X}_j) = \sum_{m=1}^{M_i} \max_n k(\mathbf{x}_{im}, \mathbf{x}_{jn}) + \sum_{n=1}^{M_j} \max_m k(\mathbf{x}_{im}, \mathbf{x}_{jn}) \quad (2.4)$$

The matching kernel is not proven to be a Mercer kernel (38; 35). The number of basic kernel computations is $2 * M_i * M_j$ for this kernel.

Construction of the summation kernel or the matching kernel is computationally intensive because each local feature vector of an example is compared with every local feature vector of the other example.

## 2.5.3 Intermediate Matching Kernel(IMK)

Assuming we have build a set of $V$ virtual features, $V = \{v_1, .....v_Q\}$, the intermediate matching kernel $K^{IMK}$ is defined as:

$$K^{IMK}(\mathbf{X}_i, \mathbf{X}_j) = \sum_{q=1}^{Q} k(\mathbf{x}_{iq}^*, \mathbf{x}_{jq}^*) \quad (2.5)$$

where $\mathbf{x}_{iq}^*$ and $\mathbf{x}_{jq}^*$ are feature vectors in $\mathbf{X}_i$ and $\mathbf{X}_j$ that are closest to virtual feature vector $\mathbf{v}_q$. More detailed explanation on IMK is included in Section 3.2.

## 2.5.4 Pyramid Match Kernel

This kernel is based on mapping a feature set to a multi-resolution histogram. In this method each feature vector is mapped to a multi-resolution histogram according to fol-

lowing function:

$$\Psi(\mathbf{X}_i) = [H_{-1}(\mathbf{X}_i), H_0(\mathbf{X}_i), ......H_L(\mathbf{X}_i)], L = \lceil log_2 D \rceil \quad (2.6)$$

where $H_j(\mathbf{X}_i)$ is a histogram of $\mathbf{X}_i$ with $d$-dimensional bins of side length $2^i$. The pyramid match kernel, $K^{PMK}$ is defined as follows:

$$\tilde{K}^{PMK}(\Psi(\mathbf{X}_i), \Psi(\mathbf{X}_j)) = \sum_{k=0}^{L} \frac{1}{2^k}(K_{HI}(H_k(\mathbf{X}_i), H_k(\mathbf{X}_j)) - K_{HI}(H_{k-1}(\mathbf{X}_i), H_{k-1}(\mathbf{X}_j))) \quad (2.7)$$

where $K_{HI}$ is histogram intersection kernel. Let $A$ and $B$ be the histogram of images $I$ and $J$ respectively then histogram kernel is defined as:

$$K_{HI}(A, B) = \sum_{i=1}^{M} min\{A_i, B_i\} \quad (2.8)$$

where $A_i$ and $B_i$ are $i^{th}$ component of histogram $A$ and $B$ respectively.

The normalization term with the bin size in Equation 2.7 represent that the matches found at finer resolution are weighted more. The difference means that if the same match appears at different resolutions, only the match found at the finer resolution is kept. To avoid favoring large input sets, a final normalization step is added:

$$K_{PMK}(\Psi(\mathbf{X}_i), \Psi(\mathbf{X}_j)) = \frac{1}{\sqrt{C}} \tilde{K}^{PMK}(\Psi(\mathbf{X}_i), \Psi(\mathbf{X}_j)), \quad (2.9)$$

where

$$C = \tilde{K}_{PMK}(\Psi(\mathbf{X}_i), \Psi(\mathbf{X}_i)) \times \tilde{K}_{PMK}(\Psi(\mathbf{X}_j), \Psi(\mathbf{X}_j)) \quad (2.10)$$

$K^{PMK}$ is proven to be mercer kernel (17).

### 2.5.5 EMD Kernel

EMD kernel is given as:

$$K_{EMD}(\mathbf{X}_i, \mathbf{X}_j) = exp(-EMD(\mathbf{X}_i, \mathbf{X}_j)/2\sigma^2) \qquad (2.11)$$

where EMD is a popular similarity measure used in image retrieval (18).

## 2.6 Performance Measure of Image Retrieval System

For measuring the performance of different image retrieval systems, we have used following two performance measures:

### 2.6.1 11-point Interpolated Average Precision

In a ranked retrieval context, appropriate sets of retrieved images are naturally given by the top k retrieved images. For each such set, precision and recall values can be plotted to give a precision-recall curve, such as the one shown in Figure 2.2

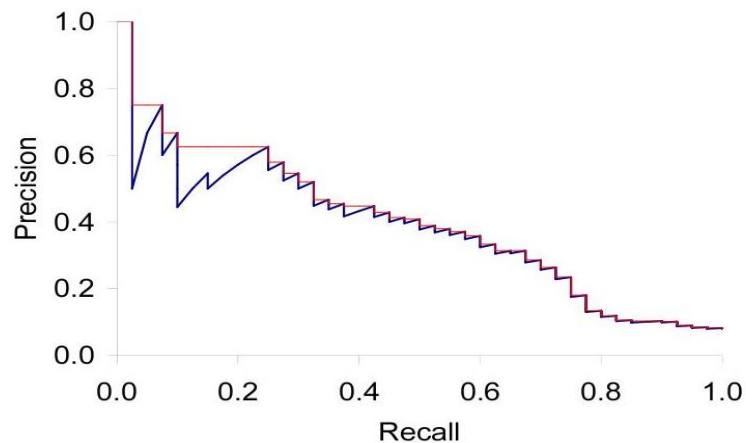

Figure 2.2: Precision/recall graph

We can see from Figure 2.2 that Precision-recall curves have a unique saw-tooth appearance, because if the $(k+1)$th image retrieved is non-relevant then recall is the same as for the top $k$ image, but precision has dropped. If it is relevant, then both precision and recall increase, and the curve shifts up and to the right. It is often useful to remove these shakes and the standard way to do this is with an interpolated precision: the interpolated precision $p_{interp}$ at a certain recall level $r$ is defined as as the highest precision found for any recall level $r' \geq r$:

$$p_{interp}(r) = max_{r' \geq r} p(r') \qquad (2.12)$$

Although, examining the entire precision-recall curve is very enlightening, but there is often a need to reduce this information down to a few numbers. The traditional way of doing this is the **11-point interpolated average precision**. For each query image, the interpolated precision is measured at the 11 recall levels of $0.0, 0.1, 0.2, ..., 1.0$. For each recall level, we then calculate the arithmetic mean of the interpolated precision at that recall level for each query image in the test collection. A composite precision- recall curve showing 11 points can then be graphed. Figure 2.3 shows an example graph of such 11-point interpolated average precision.

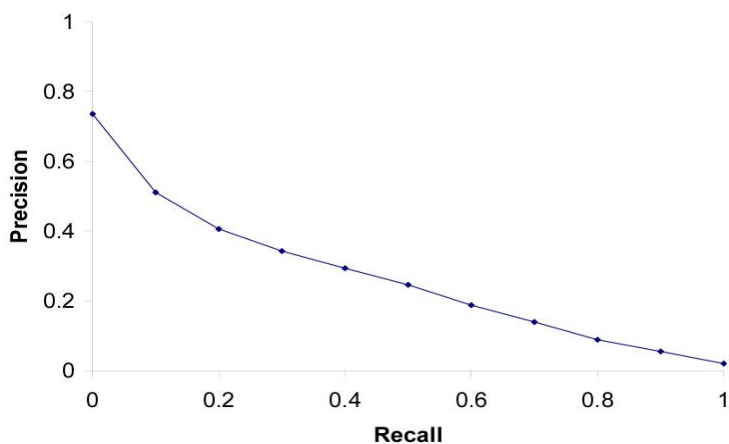

Figure 2.3: Averaged 11-point precision/recall graph across $n$ queries

## 2.6.2 Mean Average Precision

MAP provides a single-figure measure of quality of information retrieval algorithm across all recall levels. Among various evaluation measures available, MAP has been shown to have especially better separation and stability. For a single information need (in our case each query image is an individual information need), Average Precision is the average of the precision value obtained for the set of top k documents existing after each relevant document is retrieved, and this value is then averaged over all information needs. That is, if the set of relevant documents for an information need $q_j \in Q$ is $\{d_1,...d_{m_j}\}$ and $R_{jk}$ is the set of ranked retrieval results from the top result until we get to document $d_k$, then

$$MAP(Q) = \frac{1}{|Q|} \sum_{j=1}^{|Q|} \frac{1}{m_j} \sum_{k=1}^{m_j} Precision(R_{jk}) \qquad (2.13)$$

When a relevant document is not retrieved at all, the precision value in the above equation is taken to be 0. For a single information need, the average precision resemble the area under the uninterpolated precision-recall curve, and so the MAP is closely the average area under the precision-recall curve for a set of queries.

Using MAP, fixed recall levels are not considered, also, there is no interpolation. The MAP value for a set of information needs is the arithmetic mean of average precision values for individual information needs. Hence, this gives rise to a limitation of MAP that each information need is weighted equally in the final reported number, even if many documents are relevant to some queries whereas very few are relevant to other queries.

## 2.7 Performance Measure of Image Classification System

For measuring the performance of Image classification system, we have used Accuracy as measure. Accuracy is defined as follows:

$$Accuracy = \frac{Number\ of\ correctely\ classified\ test\ images}{Total\ number\ of\ test\ images} \quad (2.14)$$

## 2.8 Summary

In this chapter we have presented different learning techniques used for the task of image retrieval. Construction of image representation (visual signature) from local image feature was presented. We have also presented kernel method for set of feature vector representation. Finally, performance measure used in image retrieval and classification are presented.

# Chapter 3

# IMAGE RETRIEVAL BASED ON INTERMEDIATE MATCHING KERNEL

While designing a image retrieval systems , one of the important steps is the choice of feature representation. Global features like color histogram (31) are commonly used for image representation . Using global feature has the advantage of dealing with simple structures and thus easy algorithms can be designed for processing of features. The main drawback of such a representation is that it does not capture localized information within the images. Local representation overcomes this drawback by extracting local features. Use of local feature vector calls for more sophisticated techniques for computing similarity between a pair of images.

## 3.1 Proposed Approach for Content Based Image Retrieval using Local Feature Vectors

The goal of any CBIR technique is to compute similarity between a given query image and images in database. When images are represented using local feature vectors, then there are different ways to compute similarity between a pair of images. Most commonly, local feature vector from each image in the pair are modeled using probability distribution and similarity between these probability distributions is used as similarity between images. Also, dynamic kernels (4) can be used to compute the similarity between two images. We propose the use of Intermediate Matching Kernel, which is a dynamic kernel, for computing similarity between images (where images are represented using local feature vectors) and hence for CBIR task.

In this work, we explore two approaches for CBIR as follows:

### 3.1.1 One Step Matching

In first approach, the query image is *matched* (from matching we mean computing similarity between a pair of images) against each image in the repository. We call this approach as *One Step Matching*.

### 3.1.2 Two Step Matching

As the number of images in repository increases, the search space for each query image increases and hence the number of matchings to be performed increases. In order to reduce the search space for a query image it is necessary to identify a smaller subset of repository as potential search space. When images in repository are unsupervised (unlabeled), then a natural choice for identifying potential search spaces for different query images is using unsupervised learning algorithms like clustering. Clustering partitions whole dataset into smaller clusters ( potential search spaces in our case). For each query image, we identifying, to which cluster center it is "closest" (closeness is defined in terms of any similarity measure). Cluster corresponding to this cluster center is potential search space for the given query. After a potential search space for a query image is identified, query image is matched against images only in that search space. We call this approach as *Two Step Matching*.

We also propose to consider, for each query image, more than one nearest cluster as potential search space. intuition for considering more than one nearest cluster is as

follows:

**Intuition for considering more than one nearest cluster as potential search space**

Assigning a query point to a cluster based on distance from cluster centers is often misleading. Figure 3.1 shows cluster of data points and query points. Cluster centers are shown by + and query point are shown by ★. As we can see from figure that query point ★$_1$ belongs to cluster 1, but if we use distance from cluster centers to find the appropriate cluster for this query point, we will be get cluster 2 as matching cluster (since query point is closer to center of cluster 2 than center of cluster 1). Similarly, query point ★$_2$ belongs to cluster 3, but if we use distance from cluster centers to find the appropriate cluster for this query point, we will be get cluster 1 as matching cluster (since query point is closer to center of cluster 1 than center of cluster 2).

This leads use to conclusion that query points at the boundary of clusters are often misclassified. To overcome this problem it is appropriate to considering more than one nearest cluster as potential search space.

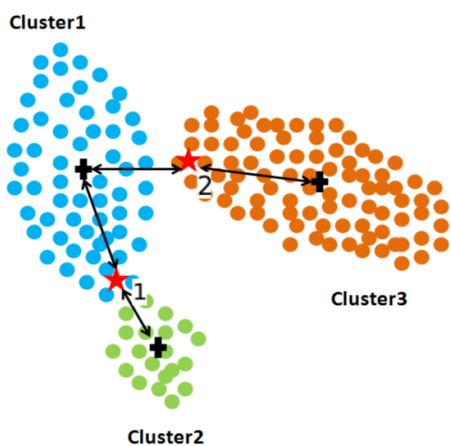

Figure 3.1: Figure showing clusters of data points and query points

Rest of the chapter is organized as follows: In Section 3.2, we present the IMK in detail. Clustering algorithms used in second step of Two Step Matching are explained in Section 3.3. Section 3.4 explains the experimental setup for One Step Matching and Two Step Matching. Finally, in Section 3.5 we present results and discussion.

## 3.2 Intermediate Matching Kernel

In this section, we present the intermediate matching kernel (IMK) that is constructed by matching the sets of local feature vectors using a set of virtual feature vectors. The construction of IMK uses a set of virtual feature vectors obtained from the training data of all the classes. The IMK for a pair of examples, with each example represented as a set of local feature vectors, is constructed by matching the local feature vectors of the examples with each of the virtual feature vectors. Consider a pair of examples $\mathbf{X}_i = \{\mathbf{x}_{i1}, \mathbf{x}_{i2}, ..., \mathbf{x}_{iM_i}\}$ and $\mathbf{X}_j = \{\mathbf{x}_{j1}, \mathbf{x}_{j2}, ..., \mathbf{x}_{jM_j}\}$ that need to be matched. Let $\mathbf{V} = \{\mathbf{v}_1, \mathbf{v}_2, ..., \mathbf{v}_Q\}$ be the set of virtual feature vectors extracted from the training data of all the classes. The feature vectors $\mathbf{x}^*_{iq}$ and $\mathbf{x}^*_{jq}$ in $\mathbf{X}_i$ and $\mathbf{X}_j$ that are closest to $\mathbf{v}_q$ are determined as follows:

$$\mathbf{x}^*_{iq} = \arg\min_{\mathbf{x} \in \mathbf{X}_i} \mathcal{D}(\mathbf{x}, \mathbf{v}_q) \quad \text{and} \quad \mathbf{x}^*_{jq} = \arg\min_{\mathbf{x} \in \mathbf{X}_j} \mathcal{D}(\mathbf{x}, \mathbf{v}_q) \tag{3.1}$$

where $\mathcal{D}(.,.)$ is a distance function that measures the distance of a local feature vector to a virtual feature vector. The process of selection of local feature vectors that are closest to the virtual feature vector is shown in Figure 3.2. Similarly, a pair of feature vectors from $\mathbf{X}_i$ and $\mathbf{X}_j$ is selected for each of the virtual feature vectors in the set. The selection of the closest local feature vectors for each virtual feature vectors involves computation of $M_i + M_j$ distance functions.

A basic kernel is computed for each of the $Q$ pairs of selected local feature vectors. The intermediate matching kernel (IMK) is computed as the sum of all the $Q$ kernel values

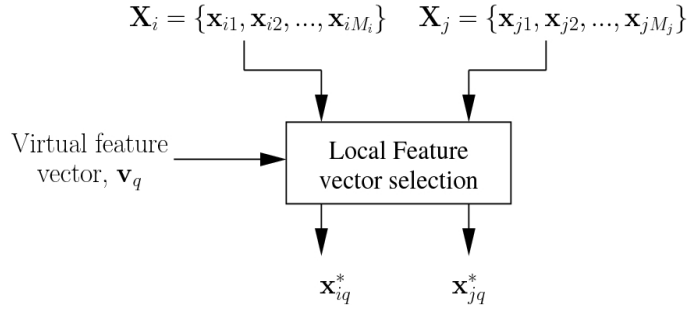

Figure 3.2: Selection of a local feature vector from each of the examples, based on the closeness to a virtual feature vector $\mathbf{v}_q$.

and is given as

$$K^{\text{IMK}}(\mathbf{X}_i, \mathbf{X}_j) = \sum_{q=1}^{Q} k(\mathbf{x}^*_{iq}, \mathbf{x}^*_{jq}) \qquad (3.2)$$

The computation of IMK involves a total of $Q*(M_i+M_j)$ computations of distance function $\mathcal{D}$ and $Q$ computations of the basic kernel. When $Q$ is significantly smaller than $M_i$ and $M_j$, the construction of IMK is computationally less intensive than constructing the summation kernel in (2.3). When $Q$ is greater than $M_i$ and $M_j$, the construction of IMK is computationally more intensive than the summation kernel in (2.3). However, it is desirable that $Q$ is smaller than the typical size of the set of local feature vectors of examples. Otherwise, a local feature vector of an example may be associated with more than one virtual feature vector.

### 3.2.1 Construction of Virtual Feature Vectors

**Set of the centers of clusters formed from the training data of all classes as the set of virtual feature vectors:**

In (35), the set of the centers of clusters formed from the training data of all classes is considered as the set of virtual feature vectors. k-means clustering method is used for clustering the training data of all the classes.

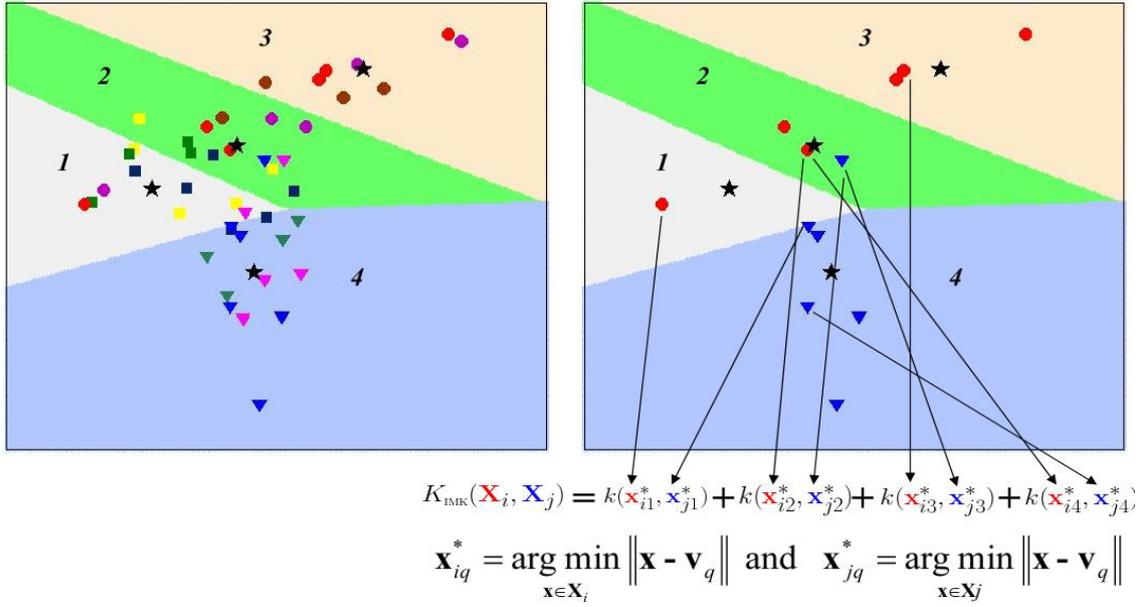

Figure 3.3: Selection of a local feature vector from each of the examples, based on the centers of clusters formed using feature vectors from all examples.

The local feature vectors $\mathbf{x}^*_{iq}$ and $\mathbf{x}^*_{jq}$ in $\mathbf{X}_i$ and $\mathbf{X}_j$ that are closest to the $q$th center $\mathbf{v}_q$ are determined as follows:

$$\mathbf{x}^*_{iq} = \arg\min_{\mathbf{x} \in \mathbf{X}_i} \|\mathbf{x} - \mathbf{v}_q\| \quad \text{and} \quad \mathbf{x}^*_{jq} = \arg\min_{\mathbf{x} \in \mathbf{X}_j} \|\mathbf{x} - \mathbf{v}_q\| \tag{3.3}$$

A basic kernel $k(\mathbf{x}^*_{iq}, \mathbf{x}^*_{jq}) = exp(-\delta \|\mathbf{x}^*_{iq} - \mathbf{x}^*_{jq}\|^2)$ is computed for each of the $Q$ pairs of selected feature vectors, where $\delta$ is a constant scaling term required for numerical stability. An IMK is computed as in Equation 3.2. Figure 3.3 explains this process.

The key issue in IMK is the selection of the set of virtual feature vectors. A better representation for the set of virtual feature vectors can be provided by considering additional information.

**Set of components of UBM as set of virtual feature vectors:**

In Dileep et.al. (36) components of the UBM built using the training data of all the classes are used for obtaining a better representation for the set of virtual feature vectors. This representation for the set of virtual feature vectors makes use of information like means of components, covariance of components and mixture coefficients. The additional information, in the form of covariance and mixture coefficients, gives a better representation for the set of virtual feature vectors as compared to the cluster centers as the set of virtual feature vectors in (35). The UBM is a large GMM of $Q$ components built using the training data of all the classes.

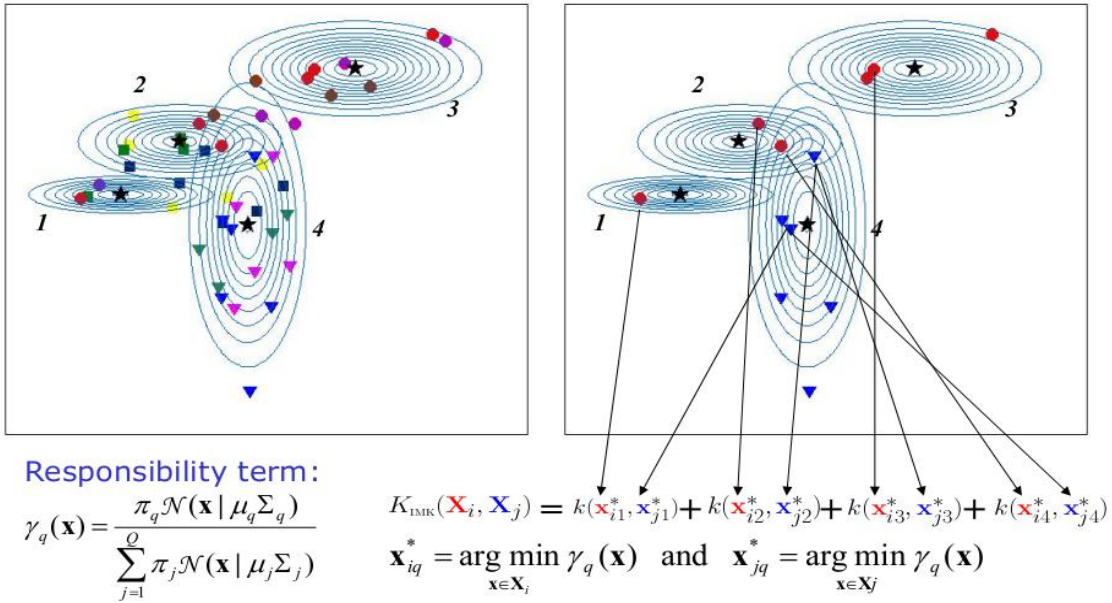

Figure 3.4: Selection of a local feature vector from each of the examples, based on UBM formed using feature vectors from all examples.

The local feature vectors from the pair of examples $\mathbf{X}_i$ and $\mathbf{X}_j$ that are closest to the component $q$ are selected for matching. Since the Gaussian components are used as the virtual feature vectors, the responsibility term is considered as a measure of closeness of the local feature vector to the component $q$. The responsibility of the component $q$ of UBM for the feature vector **x**, $\gamma_q(\mathbf{x})$, is given by

$$\gamma_q(\mathbf{x}) = \frac{\pi_q \mathcal{N}(\mathbf{x}|\boldsymbol{\mu}_q, \boldsymbol{\Sigma}_q)}{\sum_{j=1}^{Q} \pi_j \mathcal{N}(\mathbf{x}|\boldsymbol{\mu}_j, \boldsymbol{\Sigma}_j)} \quad (3.4)$$

where $\pi_q$ is the mixture coefficient of the component $q$, and $\mathcal{N}(\mathbf{x}|\boldsymbol{\mu}_q, \boldsymbol{\Sigma}_q)$ is the normal density for the component $q$ with mean $\boldsymbol{\mu}_q$ and covariance $\boldsymbol{\Sigma}_q$. The feature vectors $\mathbf{x}^*_{iq}$ and $\mathbf{x}^*_{jq}$ in $\mathbf{X}_i$ and $\mathbf{X}_j$ that are closest to the component $q$ of UBM are given by

$$\mathbf{x}^*_{iq} = \arg\max_{\mathbf{x} \in X_i} \gamma_q(\mathbf{x}) \quad \text{and} \quad \mathbf{x}^*_{jq} = \arg\max_{\mathbf{x} \in X_j} \gamma_q(\mathbf{x}) \quad (3.5)$$

A pair of local feature vectors from $\mathbf{X}_i$ and $\mathbf{X}_j$ are selected for each of the components in the UBM. The IMK is computed using a basic kernel for each of the $Q$ pairs of selected local feature vectors as in Equation 3.2. This procedure is presented in Figure 3.4.

## 3.3 Clustering Algorithms

### 3.3.1 k-means

One of the simplest and most widely used unsupervised learning algorithms that solve the clustering problem is K-means (40). This algorithm is a special case of Expectation-Maximization and follows a simple and easy way to cluster a given data set into a certain number of clusters. The number of clusters $k$ is fixed by the user. The central idea behind k-means is to define k centroids (also called as means), one for each cluster. If no prior information about centriods is available then initially random data points can be chosen as centroids . In first step, each point in the given data set is associated to its nearest centroid. This step is completed when no point is pending. After this point, we carry out second step in which we need to re-calculate k new centroids of the clusters resulting from the previous step. After we have these k new centroids, a new binding has to be done between the same data set points and the nearest new centroid. Above mentioned two steps are carried out until centroids do not change their position anymore.

Finally, this algorithm aims at minimizing an objective function given by function:

$$J = \sum_{j=1}^{k} \sum_{i=1}^{n} dist(d_i^{(j)}, m_j) \qquad (3.6)$$

where, $n$ is number of documents, $d_i^{(j)}$ is $i^{th}$ document in $j^{th}$ cluster, $m_j$ is the center of the $j^{th}$ cluster and any distance metric like Euclidean distance, Manhattan distance or kernel based dissimilarity measure can be used to computed $dist(d_i^{(j)}, m_j)$. Objective function of k-means tries to minimize intra-cluster distance among the data points.

Algorithm of k-means is given as follows :

---
**Algorithm 1** k-means Algorithm

---
**Require:** Data points $D = \{d_1, ..., d_n\}$, distance metric $dist(d_i^{(j)}, m_j)$, Number of clusters $k$.

1: Initialize random means $M = m_1, ..., m_k$.

2: **Until** there are no changes in any mean

3:     Use the estimated means to classify sample $d_1, ..., d_n$ into clusters.

4:     **For** i from 1 to k

5:         Replace $m_i$ with the mean of all the samples for cluster $i$

6:     **endfor**

7: **enduntil**

---

The results produced depend on the initial values for the means, and it frequently happens that suboptimal partitions are found. The standard solution is to try a number of different starting points.

### 3.3.2 k-medoids

The k-medoids algorithm is a clustering algorithm closely resembling the k-means algorithm. Both the k-means and k-medoids algorithms are partition based, i.e. both of them divide the dataset up into groups and both attempt to optimize some objective function that measures the quality of clusters. Simplest example of such objective function is squared error(the distance between points labeled to be in a cluster and a point designated as the center of that cluster). In opposition to the k-means algorithm, k-medoids chooses datapoints as centers (medoids or exemplars). Like k-means, k-medoid also take number of clusters from user as input. It is more robust to noise and outliers as compared to k-means because it minimizes a sum of pairwise dissimilarities instead of a sum of squared Euclidean distances. A medoid can be defined as the object of a cluster, whose average dissimilarity to all the objects in the cluster is minimal i.e. it is a most centrally located point in the cluster. Mathematically, let a cluster $c_k$ has datapoints $D_k = \{d_1, ...., d_{nk}\}$, then the medoid of the cluster $m_k$ is given as follows:

$$m_k = \arg\min_{d \in D_k} \sum_{j=1}^{d_{nk}} dist(d, d_j) \qquad (3.7)$$

k-medoids have different realizations . Most common of them is like Partition Around Medoids (PAM), CLARA (Clustering large applications), CLARANS (Clustering large applications based on randomized search). We have used simplest realization of k-medoids for our experiments. Algorithm for this realization is as follows:

**Algorithm 2** K-medoids Algorithm

**Require:** Datapoints $D = \{d_1, ..., d_n\}$, distance metric $dist(d_i, d_j)$, Number of clusters $k$.

1: Randomly select $k$ of the $n$ datapoints as the medoids $M = m_1, ..., m_k$.
2: **Until** there are no changes in any mean
3:    Associate each data point to the closest medoid.
      ("closest" here is defined using any valid distance metric, most commonly
      Euclidean distance, Manhattan distance or kernel based dissimilarity measure)
4:    **For** i from 1 to k
5:       Replace $m_i$ with the medoid of all the samples for cluster $i$
6:    **endfor**
7: **enduntil**

### 3.3.3 Affinity Propagation

Traditional data partitioning algorithm like k-centers begin with an initial set of randomly chosen exemplars for each cluster (when the centers are selected from actual data points, they are called exemplars) and iteratively improves this set so as to reduce the sum of squared errors. These clustering techniques are quite sensitive to the initial selection of exemplars, so it is usually rerun many times with different initializations in an attempt to find a good solution. However, this works well only when the number of cluster is small and chances are good that at least one random initialization is close to a good solution.

Unlike k-centers, clustered output is not dependent on initialization of exemplars in affinity propagation (AP) (37) since all the data points are assumed to be possible exemplars simultaneously. Number of clusters, $k$ is not required as a part of input. Instead, AP controls the number of identified clusters by input parameters called preferences. Pref-erence for $k^{th}$ data point denotes influence of $k^{th}$ data point to become possible exemplar. In most cases, the statistical and geometrical structure of a data set is unknown so that it is reasonable to set all the preference value the same. The bigger this shared value is, the larger the number of clusters is. Affinity propagation takes similarity values between data points as an input which is denoted by $s(i,k)$. This indicates how well data point $k$ is a possible exemplar of data point $i$.

The process of AP can be viewed as a message communication process in which each data point passes a real-valued message on the edges of the completely connected graph, where edges denotes the similarity between data points. Value of message indicates affinity i.e. weight given by one data point to select another data point as its exemplar. Two types of messages are sent across the network, availability message and responsibility message. Availability message $a(i,k)$ denotes possibility of $k^{th}$ data point to become exemplar of $i^{th}$ data point. How well $k^{th}$ data point deserves to be a possible exemplar of $i^{th}$ data point is denoted be responsibility message $r(i,k)$.

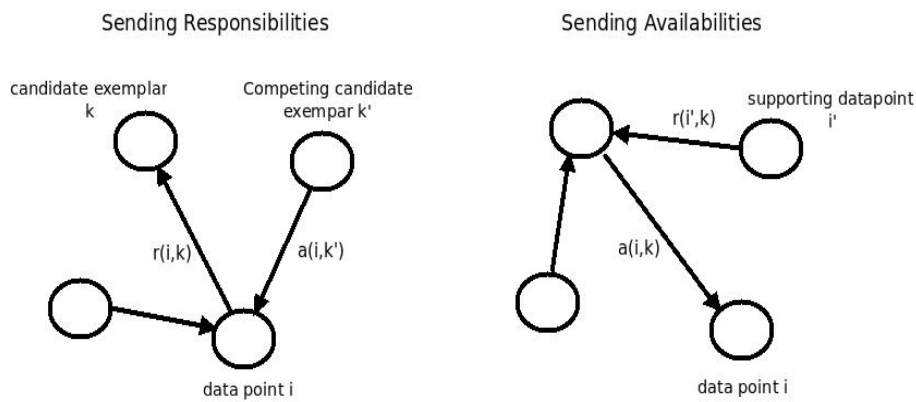

Figure 3.5: Message passing in affinity propagation (37).

Initially $a(i,k)$ is set to 0. Then, the responsibilities are computed using the rule:

$$r(i,k) \longleftarrow s(i,k) - max_{k's.tk' \neq k}\{a(i,k') + s(i,k')\} \qquad (3.8)$$

In the first iteration, because the availabilities are zero, $r(i,k)$ will be is set to the input similarity between point *i* and point *k* as its exemplar, minus the largest of the similarities between point i and other candidate exemplars. In following iterations, when some particular points are effectually assigned to other exemplars, their availabilities will drop below zero as prescribed by the update rule 3.9 . These negative availabilities will decrease the competent values of some of the input similarities s(i,k) in the above rule, removing the corresponding candidate exemplars from competition. For $k = i$, the responsibility $r(k,k)$ is set to the input preference that point *k* be chosen as an exemplar, $s(k,k)$, minus the largest of the similarities between point i and all other candidate exemplars. This self-responsibility reflects accumulated evidence that point *k* is an exemplar, based on its input preference tempered by how ill-suited it is to be assigned to another exemplar.

Whereas the above responsibility update lets all candidate exemplars the following availability update assemble evidence from data points as to whether each candidate exemplar would make a good exemplar:

$$a(i,k) \longleftarrow min\{0, r(k,k) + \sum_{i' s.t. i' \notin \{i,k\}} max\{0, r(i',k)\}\} \qquad (3.9)$$

The availability $a(i,k)$ is set to the self-responsibility $r(k,k)$ plus the sum of the positive responsibilities candidate exemplar *k* receives from other points. Only the positive portions of incoming responsibilities are added, because it is only essential for a good exemplar to explain some data points well (positive responsibilities), regardless of how poorly it explains other data points (negative responsibilities). The self-availability $a(k,k)$ is updated differently:

$$a(k,k) \longleftarrow \sum_{i' s.t. i' \notin \{i,k\}} max\{0, r(i',k)\}\} \qquad (3.10)$$

This message reflects aggregated evidence that point *k* is an exemplar, based on positive responsibilities send to candidate exemplars *k* from other points. For any point, the point which maximizes $a(i,k) + r(i,k)$ will become an exemplar for point $i$. The algorithm is terminated either after few number of iterations or changes in the availability and responsibility messages drop below certain threshold or exemplars remains constant. Formally, the algorithm for AP is described below:

---
**Algorithm 3** Affinity propagation Algorithm
---

**Require:** Datapoints $D = \{d_1, ..., d_n\}$, similarity matrix of $n$ data points, $S_{n*n}$, where the diagonal of the matrix is the preferences.

1: **Initialization:** Set availability $A_{n*n}$ to zero.
2: Updating all responsibilities $r(i,k)$:

$$r(i,k) \longleftarrow s(i,k) - max_{k's.t.k' \neq k}\{a(i,k') + s(i,k')\} \tag{3.11}$$

3: Updating all availabilities $a(i,k)$:

$$a(i,k) \longleftarrow min\{0, r(k,k) + \sum_{i's.t.i' \notin \{i,k\}} max\{0, r(i',k)\}\} \tag{3.12}$$

$$a(k,k) \longleftarrow \sum_{i's.t.i' \notin \{i,k\}} max\{0, r(i',k)\}\} \tag{3.13}$$

4: Combining availabilities and responsibilities to monitor the exemplar decisions: The data points $k$ with $a(k,k) + r(k,k) > 0$ are the identified exemplars.
5: If decisions made in step 4 did not change for a certain times of iteration or a fixed number of iteration reaches, go to step 6. Otherwise, go to step 2.
6: Assign other data points to the exemplars using the nearest assign rule, that is to assign each data point to an exemplar which it is most similar to.

---

## 3.4 Experimental Setup

### 3.4.1 Dataset

We have used UCID database. It was created as a benchmark database for CBIR and consist of a total of 1338 images of 512*384 pixels in size. There are 264 query images in the test suite with corresponding ground-truths. Out of these 264 available query images we have taken those images which have more than 3 ground truth images, as query images for our experiments.

Ground truth for query images were created by the creator of database by manual relevance assessments among all database images. The images that are judged to be relevant are images which are very clearly relevant, e.g., for an image showing a particular person, images showing the same person are searched and for an image showing a football game, images showing football games are considered to be relevant. Figure 3.6 shows example UCID images.

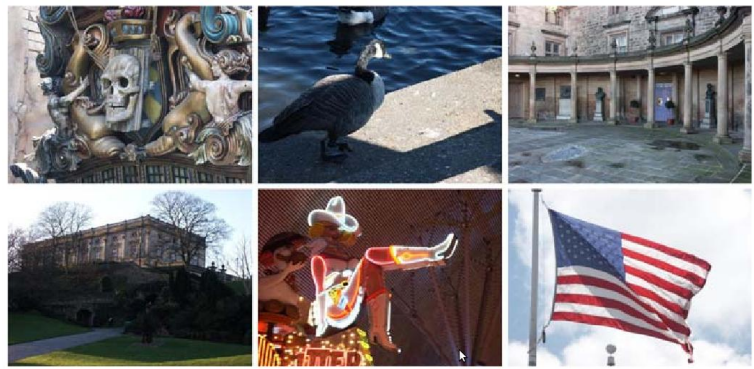

Figure 3.6: Example images from UCID database

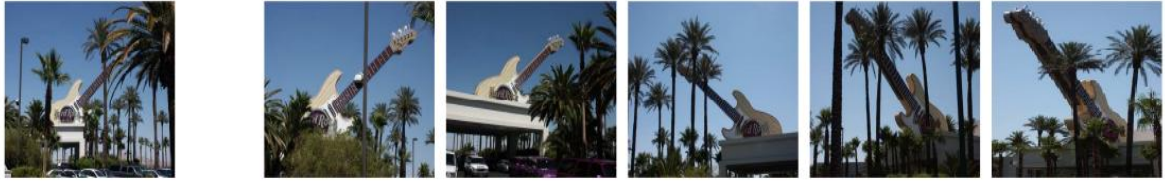

Figure 3.7: Example query image and corresponding ground truths from UCID database: Left most image is the query image and remaining are ground truth for that query image

### 3.4.2 Image Features

**Block based local feature vectors**

In block based representation, each image was partitioned into 100 equal-sized non-overlapping blocks and from each block color, edge direction histogram and texture feature were extracted. For color features, three moments per color channel in HSV color space were extracted, accounting for 9 elements in the feature vector. For the normalized edge direction histogram features, 8 bins of 45 degrees each were considered. This leads to 8 elements in the feature vector. For texture feature, wavelet decomposition of each block up to two level was used. From each of the decomposed blocks, entropy feature was extracted leading to 6 texture features. Thus, a 23-dimensional feature vector was used to represent a block of an image. In this way each image was represented by 100 23-dimensional feature vectors.

**Salient point based local feature vectors**

In salient point representation of image, we have first extracted salient point from images using salient point extraction algorithm (45). After this, we have taken block of size 51*39 surrounding each salient point and extracted 23-dimensional feature vector from those blocks (23-dimensional feature vector is same as described above). Since the number of salient point vary from image to image, so images are represented as varying number of local feature vectors, i.e., in our representation each image consist of 89 to 127 23-dimensional local feature vector.

### 3.4.3 Baseline Image Retrieval System

In our experiments, we have used Gaussian mixture model based image retrieval system as baseline. Gaussian mixtures (GM) are a well-established methodology to model probability density functions (pdf). The advantages of this methodology, such as adaptability to the data, modeling flexibility and robustness, have made GM models attractive for a wide range of applications. Because of the above merits, GM models have already been employed for the CBIR problem(41; 42; 43). The main challenge when using a GM model in CBIR is to define a distance measure between GMs, which separates the different models well and, in addition, can be computed efficiently. The traditionally used distance measure between pdfs is the Kullback Leibler (KL) divergence. For two pdfs, $p_1(x)$ and $p_2(x)$, is defined as:

$$KL(p_1\|p_2) = \int p_1(\mathbf{x}) \frac{p_1(\mathbf{x})}{p_2(\mathbf{x})} d\mathbf{x} \qquad (3.14)$$

KL divergence can be computed in closed form for Gaussian pdf's but it cannot be computed in closed form for GM models. To overcome this limitation a new distance measure $C2$ was proposed in (44) that can be computed in closed form for GM models. This measure between two pdfs, $p_1(\mathbf{x})$ and $p_2(\mathbf{x})$, is defined as:

$$C2(p_1, p_2) = -\log \frac{2S_{p_1 p_2}}{S_{p_1 p_1} + S_{p_2 p_2}} \qquad (3.15)$$

with

$$S_{p_m p_l} = S_{ml} = \int p_m(\mathbf{x}) p_l(\mathbf{x}) dx \qquad (3.16)$$

when $p_1(\mathbf{x})$ and $p_2(\mathbf{x})$ are GMs, we have

$$S_{ml} = \frac{1}{\sqrt{(2\pi)^d}} \sum_{i=1}^{k_m} \sum_{j=1}^{k_l} \frac{\pi_{mi}\pi_{li}}{\sqrt{|C_{ml}(i,j)|}e^{k_{ml}(i,j)}} \qquad (3.17)$$

where

$C_{ml}(i,j) = \Sigma_{mi} + \Sigma_{lj}$

$k_{ml}(i,j) = (\mu_{mi} - \mu lj)^T C_{ml}^{-1}(i,j)(\mu_{mi} - \mu lj)$

In Section 3.5 we have used following notation for baseline systems:

- KLD : A retrieval system in which each image is modeled using Gaussian pdf and image similarity is computed using KL divergence between pdf's.

- C2-1 : A retrieval system in which each image is modeled using Gaussian pdf and image similarity is computed using C2 divergence between pdf's.

- C2-2: A retrieval system in which each image is modeled using two component GMM and image similarity is computed using C2 divergence between GMM's.

## 3.5 Results and Discussions

### 3.5.1 IMK Based One-step Matching for Image Retrieval

Table 3.1 presents the MAP value obtained using different baseline systems and IMK based system. We observe that for both type of local feature vectors, when components of UBM are used in feature vector selection in IMK system ,the system performs significantly(21%− 40%) better than all other systems. When center of clusters are used for feature vector selection in IMK based system, the performance of system is 7% − 8% less than KLD

| Local Feature Vectors | Mean Average Precision (MAP) values | | | | |
|---|---|---|---|---|---|
| | Distribution Based | | | Kernel Based(IMK) | |
| | KLD | C2-1 | C2-2 | Centers of Clusters | Components of UBM |
| Block based | .2614 | .2693 | .1952 | .2406 | **.3369** |
| Salient point based | .2186 | .2641 | .1759 | .2409 | **.3315** |

Table 3.1: Comparison of different image retrieval systems using Mean Average Precision as a performance measure for One Step Matching

and C2-1 system (except for KLD system that uses salient point based representation of images).

Figure 3.8 shows the 11-point interpolated average precision curve for different retrieval systems using both type of local feature vectors. These curves show that precision of Components of UBM based IMK system is better than all other systems, across all values of recall.

Finally, we can summarize the performance of different systems as follows:

*C2–2 < IMK–Center of Clusters < KLD ≈ C2–1 < IMK–Components of UBM*

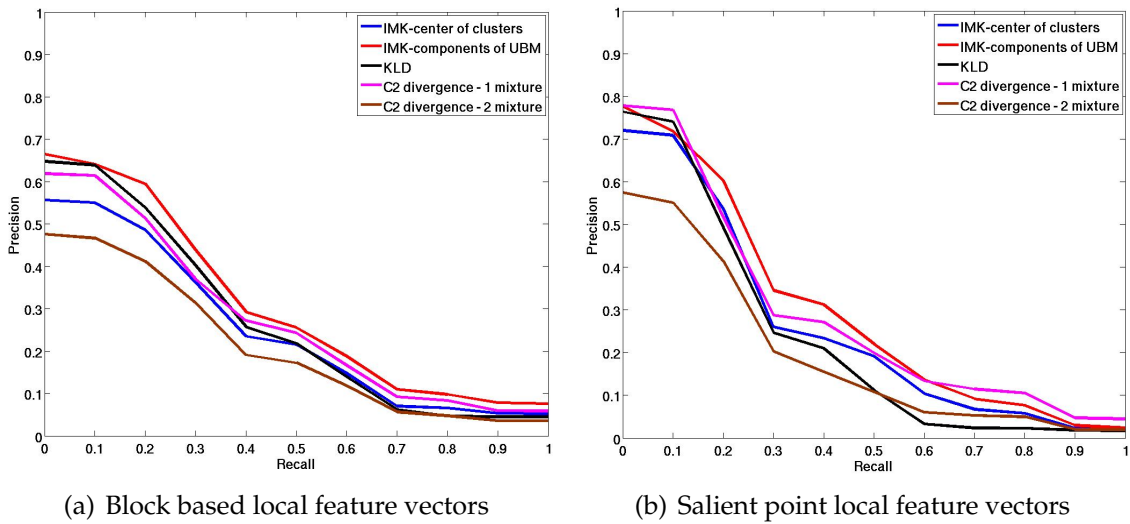

(a) Block based local feature vectors  (b) Salient point local feature vectors

Figure 3.8: Comparison of different image retrieval systems using 11-point interpolated Average Precision as performance measure for One Step Matching

### 3.5.2 IMK Based Two-Step Matching for Image Retrieval

The aim of two step matching is reduction in search space without suffering from significant drop in retrieval performance. In two step matching, first issue is the choice of clustering algorithm, since final performance depends critically on clustering step. When block-based local feature vectors of images are used, then one natural choice of clustering algorithm is using k-means algorithm, with euclidean distance between super-vector representation of images as similarity measure. Table 3.2 shows results obtained for block based local feature vectors, using k-means clustering in first step and IMK based matching in second step. Like One step matching, here also we found that UBM based IMK is performing better than cluster based IMK with significant margin. Thus, for further studies we have considered only UBM based IMK.

Choosing more than one nearest clusters for matching in second step gives much better performance than choosing only single nearest cluster. This justifies our intuition made in Section 3.1.2. If we compare the performance of two step matching with one step matching, we find that there is substantial reduction in search space with out incurring much loss in MAP value.

Now, instead of using euclidean distance, we use IMK as a similarity measure in clustering algorithms. Table 3.3 and Table 3.4 present two step matching performance obtained using IMK based k-medoid and IMK based Affinity propagation algorithm for block based local feature vectors and salient point based local feature vectors respectively. We note that, for either cases when IMK based k-medoid or IMK based Affinity propagation was used as clustering algorithm in first step, reduction in search space is almost same, across all values of K (number of clusters formed in first step), but the MAP value increases if we use IMK based Affinity propagation clustering instead of IMK based k-medoid clustering in first step.

| Mean Average Precision (MAP) values | | | | |
|---|---|---|---|---|
| Number of clusters formed in first step | Number of clusters chosen for matching in second step | % Reduction in search space | Selection of virtual feature vector for IMK based matching using | |
| | | | UBM | center of clusters |
| K=5 | 1 | 78.8 | .2533 | .1961 |
| K=10 | 1 | 87.7 | .2189 | .1802 |
| K=15 | 1 | 95.5 | .1785 | .1582 |
| K=20 | 1 | 93.4 | .1874 | .1736 |
| K=25 | 1 | 95.2 | .1716 | .1500 |
| K=5 | 3 | 37.6 | .3205 | .2386 |
| K=10 | 3 | 68.1 | .2905 | .2483 |
| K=15 | 3 | 77.2 | .2560 | .2218 |
| K=20 | 3 | 82.4 | .2769 | .2294 |
| K=25 | 3 | 85.7 | .2459 | .2058 |

Table 3.2: Two Step Matching MAP Performance for Block based local feature vectors : First Step= Clustering using k-means on super-vector (Euclidean distance as dissimilarity measure) , Second Step= Matching using IMK

| Number of clusters formed in first step | Number of clusters chosen for matching in second step | Clustering technique used in first step | | | |
|---|---|---|---|---|---|
| | | k-medoid using IMK as similarity measure | | AP using IMK as similarity measure | |
| | | % Reduction in search space | MAP | % Reduction in search space | MAP |
| K=5 | 1 | 75.5 | .2679 | 76.5 | .2931 |
| K=10 | 1 | 88.1 | .2256 | 88 | .2589 |
| K=15 | 1 | 91.9 | .2066 | 90.9 | .2057 |
| K=20 | 1 | 93.9 | .1821 | 93.9 | .1898 |
| K=25 | 1 | 94.8 | .1781 | 94.9 | .1979 |
| K=5 | 3 | 37.5 | .3266 | 30 | .3358 |
| K=10 | 3 | 66.3 | .2915 | 66.6 | .3290 |
| K=15 | 3 | 76.6 | .2914 | 76.9 | .3141 |
| K=20 | 3 | 81.5 | .2887 | 82.3 | .2903 |
| K=25 | 3 | 85.1 | .2722 | 85.2 | .2898 |

Table 3.3: Two Step Matching MAP Performance for Block based local feature vectors : First Step= Clustering using k-medoids /AP (using IMK as similarity measure) , Second Step= Matching using IMK

| Number of clusters formed in first step | Number of clusters chosen for matching in second step | Clustering technique used in first step | | | |
|---|---|---|---|---|---|
| | | k-medoid using IMK as similarity measure | | AP using IMK as similarity measure | |
| | | % Reduction in search space | MAP | % Reduction in search space | MAP |
| K=5 | 1 | 74.5 | .2234 | 76.8 | .2729 |
| K=10 | 1 | 89.3 | .2074 | 90.8 | .2348 |
| K=15 | 1 | 92.8 | .2065 | 82.2 | .2303 |
| K=20 | 1 | 93.9 | .1892 | 93.7 | .2265 |
| K=25 | 1 | 95 | .1734 | 95 | .2190 |
| K=5 | 3 | 37.6 | .3211 | 37.3 | .3280 |
| K=10 | 3 | 66.3 | .2992 | 68 | .3017 |
| K=15 | 3 | 77.3 | .3010 | 77.2 | .3011 |
| K=20 | 3 | 83 | .2878 | 82.4 | .2935 |
| K=25 | 3 | 86.3 | .2773 | 85.7 | .2930 |

Table 3.4: Two Step Matching MAP Performance on Salient point based local feature vectors : First Step= Clustering using k-medoids /AP (using IMK as similarity measure) , Second Step= Matching using IMK

Finally, we evaluate the effect of clustering algorithm used in first step on retrieval performance of two step matching system. Figure 3.9 and Figure 3.10 depict the performance of two step matching system with respect to different underlying clustering algorithms used in first step. It is observable that use of Affinity propagation as a clustering algorithm results in better retrieval performance.

## 3.6 Summary

In this chapter we have evaluated the performance of IMK based image retrieval system. We found that it performs notably better than baseline systems. In context of two step matching, we note that choosing more than one clusters for matching instead of choosing single cluster, results in remarkable performance improvement.

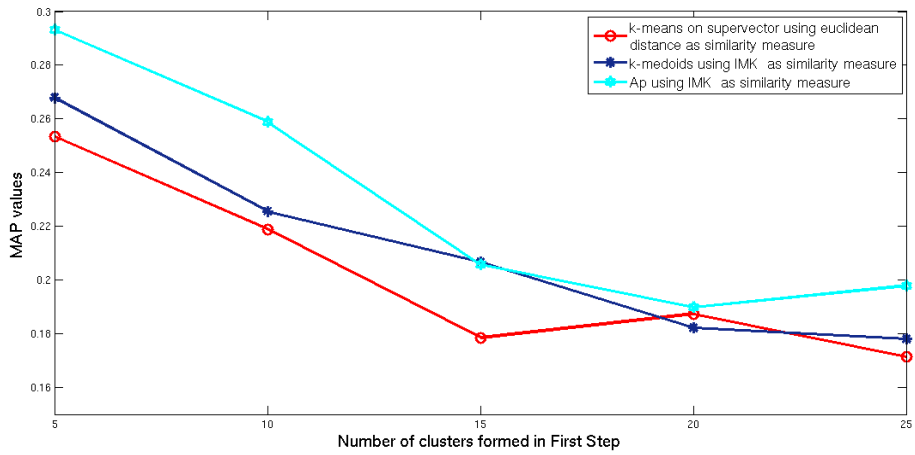

(a) Number of cluster chosen for matching in Second step=1

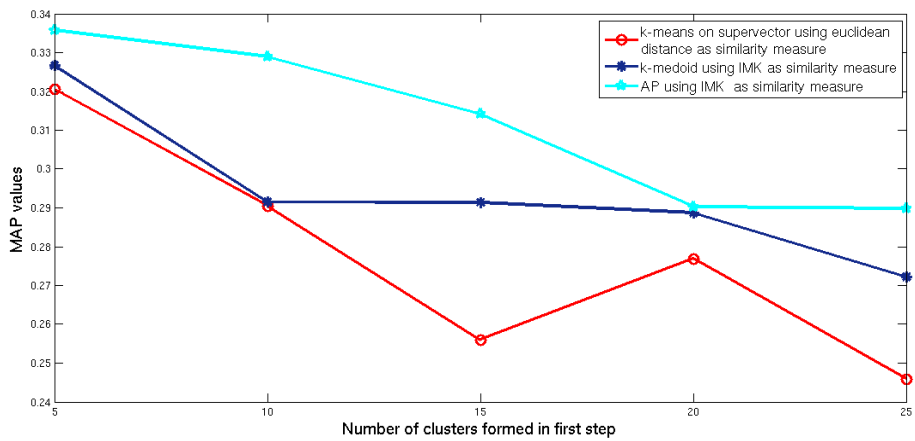

(b) Number of cluster chosen for matching in Second step=3

Figure 3.9: Effect of different clustering algorithms(used in First step) on performance of Two step IMK based image retrieval system for Block based local feature vectors

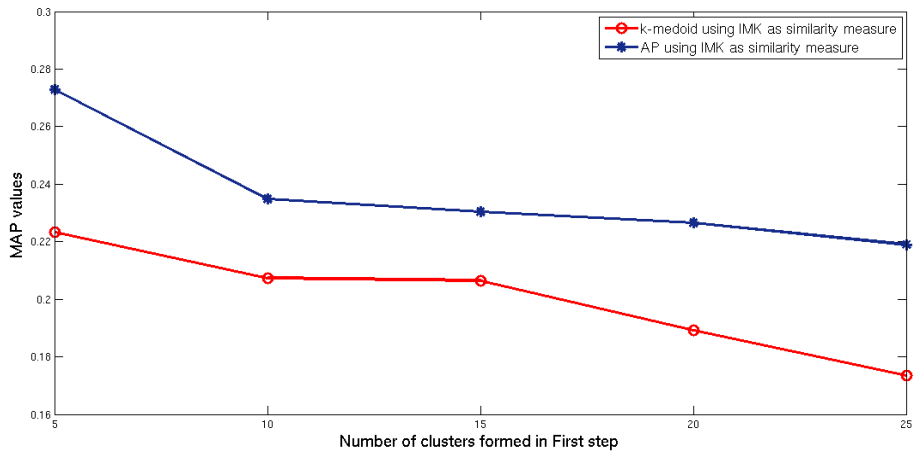

(a) Number of cluster chosen for matching in Second step=1

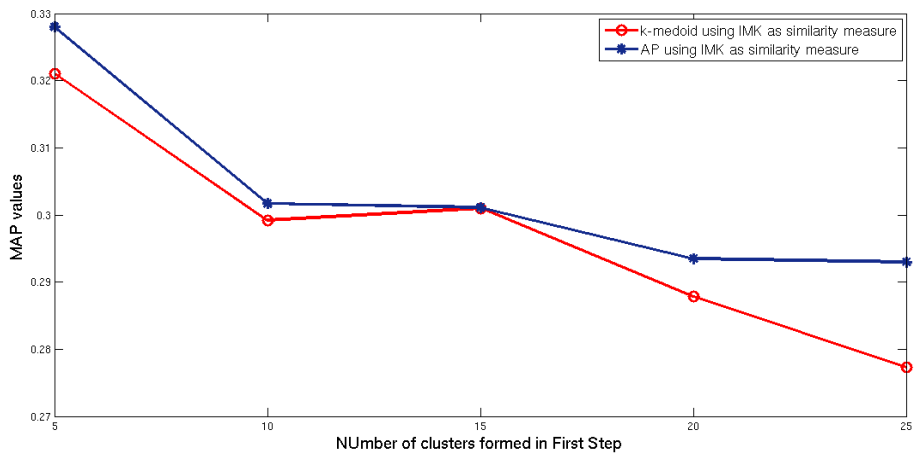

(b) Number of cluster chosen for matching in Second step=3

Figure 3.10: Effect of different clustering algorithms(used in First step) on performance of Two step IMK based image retrieval system for Salient point based local feature vectors

# Chapter 4

# A META-LEARNING FRAMEWORK FOR IMAGE RETRIEVAL

In previous chapter we investigated the performance of different Image Retrieval systems. We have realized that even the sophisticated systems like IMK based image retrieval system give only modest performance guarantees. To improve the performance of existing image retrieval systems, we have developed a meta-learning framework and investigated whether this framework gives better performance than underlying image retrieval systems. This framework is inspired by Datta et. al. (46). The framework includes a probabilistic approach to meta-learning. This probabilistic approach makes use of all accessible information, including underlying 'black-box' performance, ground-truth available for some images and alternate image representation.

Given a image retrieval system or algorithm, we employ it as 'black-box 'and develop a meta-learner that tries to conceive the performance of image retrieval system on each of the images in system's search space, using all accessible information, which includes:

- retrieval result of underlying black-box.
- Ground-truth for images, whenever accessible.
- Alternate representation of images.

We are concerned only about the output of black-box, rather being concerned directly with the mechanism black-box uses. Also, this framework does not make use of ranked ordering of the images retrieved by black-box, instead it treats them as a collection of images.

Ground-truth may be readily available for a subset of images, or in an user-feedback setting, where user judges the relevance of retrieved images and sends feedback to system, it is made available as and when user sends feedback. Figure 4.1 depicts the meta-learning framework.

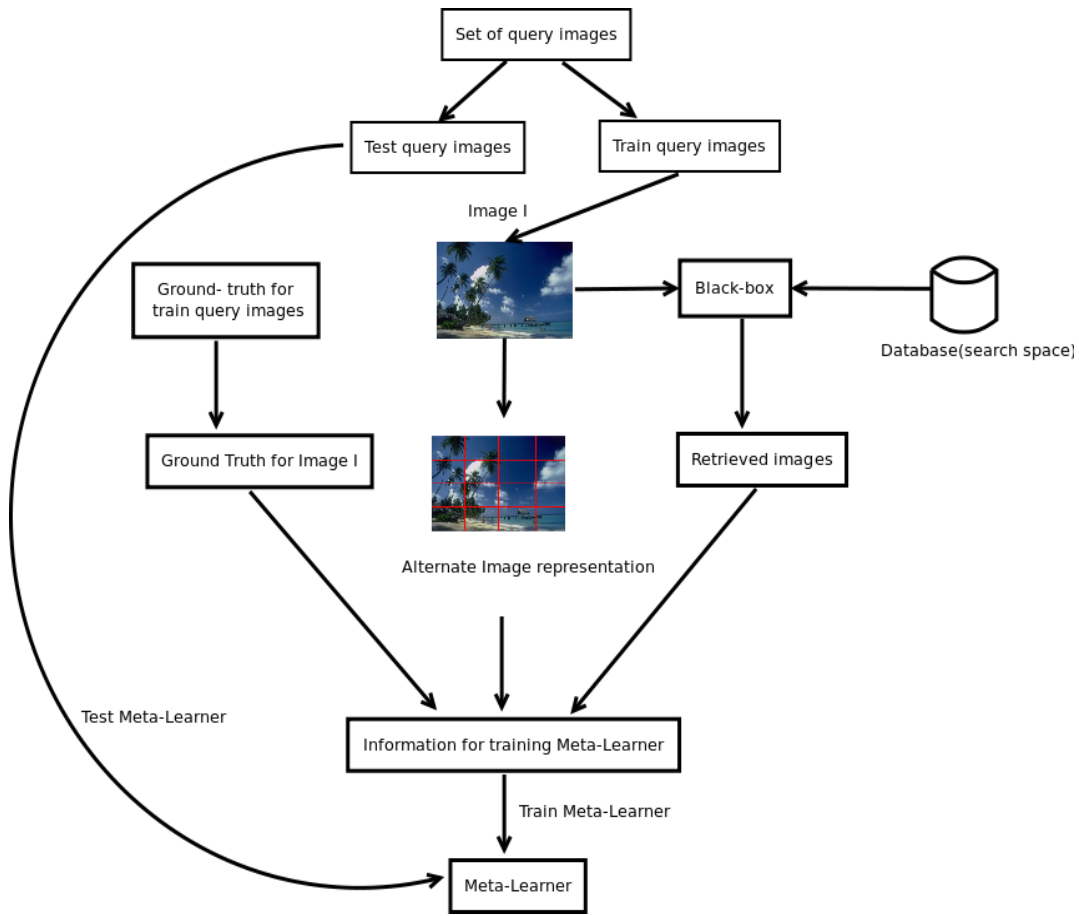

Figure 4.1: Meta-Learning framework for image retrieval

## 4.1 Generic Framework

Let the search space for a given query image $I$ be denoted by $S = \{I_1, I_2, .....I_K\}$. Given an query image $I$, the black-box predicts set of images to be its semantically similar images. To denote these retrieval results, we introduce indicator variables $R_{I_j} \in \{0, 1\}, I_j \in S$, where a value of 1 or 0 indicates whether image $I_j$ is retrieved by the black-box for $I$ or not. We define, similar indicator variables $G_{I_j} \in \{0, 1\}, I_j \in S$ to denote the ground-truth corresponding to given query $I$ image, where a value of 1 or 0 indicates whether $I_j$ is a ground-truth for $I$ or not. In a sense, we are representing the black-box as a multi-valued function $f_{bbox}$ mapping an query image $I$ to indicator variables $R_{I_j}$, i.e. $f_{bbox}(I) = (R_{I_1}, ..........., R_{I_K})$. Similarly, the function $f_{gtruth}(I)$ is a mapping from query image $I$ to its ground-truth images using the indicator variables $G_{I_j}$ i.e. $f_{gtruth}(I) = (G_{I_1}, ..........., G_{I_K})$.

Apart from this, regardless of image-representation used by black-box for image retrieval, the pixel-level image representation may be nevertheless available to the meta-learner. If some visual features extracted from the images form a different representation than what the black-box uses, they can be thought of as a different viewpoint and thus be potentially useful for semantics recognition. Such a feature representation should be easy enough so that it does not add compelling computational overhead. This visual representation can be conceived of as a function defined as: $f_{vis}(I) = (h_1, ..., h_D)$. Here, we specify a $D$-dimensional image feature vector representation as an example. Alternatively, other representations (e.g., variable-length region-based features) can also be used as long as they are efficient to compute and process, so as to not to incur significant time for training meta-learner.

Now we present the probabilistic formulation of meta-learner. In principle, this meta-learner, trained on available data along with accessible ground-truth (see Figure 4.1), acts a function which takes in all available information associated to an image $I$, including the black-boxs retrieval result, and produces a new ordered set of images as its retrieval result. In our meta-learner, this function is realized by taking decisions on each image in search space independently. For doing so, we compute the following "*odds*" in favor of each image $I_j$ to be an actual ground-truth image for image $I$, given all useful information,

as follows:
$$\ell_{I_j}(I) = \frac{Pr(G_{I_j} = 1|f_{bbox}(I), f_{vis}(I))}{Pr(G_{I_j} = 0|f_{bbox}(I), f_{vis}(I))} \tag{4.1}$$

Using Bayes Rule, we can re-write Eq. 4.1:

$$\ell_{I_j}(I) = \frac{Pr(G_{I_j} = 1, f_{bbox}(I), f_{vis}(I))}{Pr(f_{bbox}(I), f_{vis}(I))} \times \frac{Pr(f_{bbox}(I), f_{vis}(I))}{Pr(G_{I_j} = 0, f_{bbox}(I), f_{vis}(I))} \tag{4.2}$$

$$\ell_{I_j}(I) = \frac{Pr(G_{I_j} = 1, f_{bbox}(I), f_{vis}(I))}{Pr(G_{I_j} = 0, f_{bbox}(I), f_{vis}(I))} \tag{4.3}$$

In $f_{bbox}$, if the realization of variable $R_{I_i}$ for each image $I_i$ is denoted by $r_i \in \{0,1\}$ given $I$, then without loss of generality, for each $j$, we can split $f_{bbox}$ as follows:

$$f_{bbox}(I) = (R_{I_j} = r_j, \bigcup_{i \neq j}(R_{I_i} = r_i)) \tag{4.4}$$

We now evaluate the joint probability in the numerator and denominator of $\ell_{I_j}$ separately, using Eq. 4.4. For a realization $g_j \in \{0,1\}$ of the random variable $G_{I_j}$, we can factor the joint probability (using the chain rule of probability) into a prior and series of conditional probabilities, as follows:

$$\begin{aligned} Pr(G_{I_j} = g_j, f_{bbox}(I), f_{vis}(I)) = &\; Pr(R_{I_j} = r_j) \times Pr(G_{I_j} = g_j \mid R_{I_j} = r_j) \\ &\times Pr(\bigcup_{i \neq j}(R_{I_i} = r_i) \mid G_{I_j} = g_j, R_{I_j} = r_j) \\ &\times Pr(f_{vis}(I) \mid \bigcup_{i \neq j}(R_{I_i} = r_i), G_{I_j} = g_j, R_{I_j} = r_j) \end{aligned} \tag{4.5}$$

The odds in Eq. 4.1 can now be factored using Eq. 4.3 and 4.5:

$$\ell_{I_j}(I) = \frac{Pr(G_{I_j} = 1 \mid R_{I_j} = r_j)}{Pr(G_{I_j} = 0 \mid R_{I_j} = r_j)} \\ \times \frac{Pr(\bigcup_{i \neq j}(R_{I_i} = r_i) \mid G_{I_j} = 1, R_{I_j} = r_j)}{Pr(\bigcup_{i \neq j}(R_{I_i} = r_i) \mid G_{I_j} = 0, R_{I_j} = r_j)} \\ \times \frac{Pr(f_{vis}(I) \mid G_{I_j} = 1, \bigcup_{i \neq j}(R_{I_i} = r_i), R_{I_j} = r_j)}{Pr(f_{vis}(I) \mid G_{I_j} = 0, \bigcup_{i \neq j}(R_{I_i} = r_i), R_{I_j} = r_j)} \quad (4.6)$$

The ratio $\frac{Pr(R_{I_j}=r_j)}{Pr(R_{I_j}=r_j)} = 1$, and hence is eliminated. The ratio $\frac{Pr(G_{I_j}=1 \mid R_{I_j}=r_j)}{Pr(G_{I_j}=0 \mid R_{I_j}=r_j)}$ is a *performance check* on black-box for each image in search space. For $R_{I_j} = 1$, it can be interpreted as "Given that image $I_j$ is retrieved by the black-box for query image $I$, what are the odds of it being semantically similar to query image?". A higher odds indicates that the black-box has greater precision in retrieval (i.e., when $I_j$ is retrieved, it is usually semantically similar to query image). A similar interpretation can be done for $R_{I_j} = 0$, where higher odds implies lower image-specific recall in the black box guesses. These probability ratios therefore give the meta-learner indications about the black-box models performance for each image in the search space.

The term $\frac{Pr(\bigcup_{i \neq j}(R_{I_i}=r_i) \mid G_{I_j}=1,R_{I_j}=r_j)}{Pr(\bigcup_{i \neq j}(R_{I_i}=r_i) \mid G_{I_j}=0,R_{I_j}=r_j)}$ in Eq. 4.6 models the retrieval of image $I_j$ with every other image $I_i$, $i \neq j$, given that other image $I_j$ is correctly/wrongly retrieved. Since the meta-learner makes decisions about each image independently, we can separate them out in this ratio as well. That is, the retrieval of each image $I_i$ is conditionally independent of each other, given a correctly/wrongly retrieved image $I_j$, leading to the following approximation:

$$Pr(\bigcup_{i \neq j}(R_{I_i} = r_i) \mid G_{I_j} = g_j, R_{I_j} = r_j) \approx \prod_{i \neq j} Pr(R_{I_i} = r_i \mid G_{I_j} = g_j, R_{I_j} = r_j) \quad (4.7)$$

The ratio can then be written as:

$$\frac{Pr(\bigcup_{i \neq j}(R_{I_i} = r_i) \mid G_{I_j} = 1, R_{I_j} = r_j)}{Pr(\bigcup_{i \neq j}(R_{I_i} = r_i) \mid G_{I_j} = 0, R_{I_j} = r_j)} = \prod_{i \neq j} \frac{Pr(R_{I_i} = r_i \mid G_{I_j} = 1, R_{I_j} = r_j)}{Pr(R_{I_i} = r_i \mid G_{I_j} = 0, R_{I_j} = r_j)} \quad (4.8)$$

Finally, $\frac{Pr(f_{vis}(I) \mid G_{I_j}=1, \bigcup_{i \neq j}(R_{I_i}=r_i), R_{I_j}=r_j)}{Pr(f_{vis}(I) \mid G_{I_j}=0, \bigcup_{i \neq j}(R_{I_i}=r_i), R_{I_j}=r_j)}$ can be simplified, since $f_{vis}(I)$ is meta-learner's own visual representation unassociated to the black-box's visual abstraction used for retrieving images. Therefore, we can write:

$$\frac{Pr(f_{vis}(I) \mid G_{I_j} = 1, \bigcup_{i \neq j}(R_{I_i} = r_i), R_{I_j} = r_j)}{Pr(f_{vis}(I) \mid G_{I_j} = 0, \bigcup_{i \neq j}(R_{I_i} = r_i), R_{I_j} = r_j)} \approx \frac{Pr(h_1, \ldots, h_D \mid G_{I_j} = 1)}{Pr(h_1, \ldots, h_D \mid G_{I_j} = 0)} \quad (4.9)$$

Putting everything together, and taking logarithm to get around issues of machine precision, we can re-write Eq. 4.6 as:

$$\begin{aligned}
\log \ell_{I_j}(I) = & \log \left( \frac{Pr(G_{I_j} = 1 \mid R_{I_j} = r_j)}{1 - Pr(G_{I_j} = 1 \mid R_{I_j} = r_j)} \right) \\
& + \sum_{i \neq j} \log \left( \frac{Pr(R_{I_i} = r_i \mid G_{I_j} = 1, R_{I_j} = r_j)}{Pr(R_{I_i} = r_i \mid G_{I_j} = 0, R_{I_j} = r_j)} \right) \\
& + \log \left( \frac{Pr(h_1, \ldots, h_D \mid G_{I_j} = 1)}{Pr(h_1, \ldots, h_D \mid G_{I_j} = 0)} \right)
\end{aligned} \quad (4.10)$$

## 4.2 Parameter Estimation and Smoothing

The core of the meta-learner is Eq. 4.10, which takes in an query image $I$ and the black-box retrieval for it, and subsequently computes odds for each image in search space. The probabilities are assessed from any data available to meta-learner that consist of a set of images, associated ground-truths and retrieval results obtained through black-box. Next we present the estimation of each term separately, given a *training data* of size $L$, consisting of images $\{I^{(1)}, \ldots, I^{(L)}\}$, the corresponding retrieval made by the black-box $\{f_{bbox}(I^{(1)}), \ldots f_{bbox}(I^{(L)})\}$ and the actual ground-truth $\{f_{gtruth}(I^{(1)}), \ldots f_{gtruth}(I^{(L)})\}$.

The probability $Pr(G_{I_j} = 1 \mid R_{I_j} = r_j)$ in Eq. 4.10 can be estimated from the size $L$ training data as follows:

$$\hat{Pr}(G_{I_j} = 1 \mid R_{I_j} = r_j) = \frac{\sum_{n=1}^{L} \mathcal{I}\{R_{I_j}^{(n)} = r_j \ \& \ G_{I_j}^{(n)} = 1\}}{\sum_{n=1}^{L} \mathcal{I}\{R_{I_j}^{(n)} = r_j\}} \qquad (4.11)$$

Here, $\mathcal{I}(.)$ is the indicator function. When the training set contains too few or no samples for $R_{I_j} = 1$, then estimation will be poor or impossible. Therefore, we a standard *interpolation- based smoothing* of probabilities is performed. For this we require a *prior* estimate, which computed as follows:

$$\hat{Pr}_{prior}(r) = \frac{\sum_{i=1}^{K} \sum_{n=1}^{L} \mathcal{I}\{R_{I_j}^{(n)} = r \ \& \ G_{I_j}^{(n)} = 1\}}{\sum_{i=1}^{K} \sum_{n=1}^{L} \mathcal{I}\{R_{I_j}^{(n)} = r\}} \qquad (4.12)$$

where $r \in \{0, 1\}$. The image-specific estimates are interpolated with the prior to get the final estimates as follows:

$$\tilde{Pr}(G_{I_j} = 1 \mid R_{I_j} = r_j) = \begin{cases} \hat{Pr}_{prior}(r_j) & m \leq 1 \\ \frac{1}{m}\hat{Pr}_{prior}(r_j) + \frac{m}{m+1}\hat{Pr}(G_{I_j} = 1 \mid R_{I_j} = r_j) & m > 1 \end{cases} \qquad (4.13)$$

where $m = \sum_{n=1}^{L} \mathcal{I}\{R_{I_j}^{(n)} = r_j\}$, the number of instances out of $L$ where $I_j$ was retrieved (or not retrieved depending upon $r_j$).

The probability $Pr(R_{I_i} = r_i \mid G_{I_j} = 1, R_{I_j} = r_j)$ in Eq. 4.10 can be estimated from the training data as follows:

$$\hat{Pr}(R_{I_i} = r_i \mid G_{I_j} = 1, R_{I_j} = r_j) = \frac{\sum_{n=1}^{L} \mathcal{I}\{R_{I_i} = r_i \ \& \ R_{I_j}^{(n)} = r_j \ \& \ G_{I_j}^{(n)} = 1\}}{\sum_{n=1}^{L} \mathcal{I}\{R_{I_j}^{(n)} = r_j \ \& \ G_{I_j}^{(n)} = 1\}} \qquad (4.14)$$

Similarly, the probability $Pr(R_{I_i} = r_i \mid G_{I_j} = 0, R_{I_j} = r_j)$ in Eq. 4.10 can be estimated from the training data as follows:

$$\hat{Pr}(R_{I_i} = r_i \mid G_{I_j} = 0, R_{I_j} = r_j) = \frac{\sum_{n=1}^{L} \mathcal{I}\{R_{I_i} = r_i \ \& \ R_{I_j}^{(n)} = r_j \ \& \ G_{I_j}^{(n)} = 0\}}{\sum_{n=1}^{L} \mathcal{I}\{R_{I_j}^{(n)} = r_j \ \& \ G_{I_j}^{(n)} = 0\}} \quad (4.15)$$

Finally, we discuss component $Pr(h_1, \ldots, h_D \mid G_{I_j} = g), g \in \{0, 1\}$, in Eq. 4.10. The idea is that the probabilistic model for a particular visual representation may differ when a certain word is correct, versus when it is not. Among various available feature representation, we choose feature representation that is easily computed so as not to incur compelling computation cost in meta-learning framework. Each image is divided into 16 equal parts by cutting along each axis into four equal parts. Within each block, L, a, b space values for each pixel were computed and the triplet of average L, a, b value represent that block. These triplets were concatenated in raster order of the blocks from top-left to obtain 48-dimensional vector $(h_1 \ldots h_{48})$. For estimation each of the 48 components fitted with a univariate Gaussian, which involves calculating the estimated mean $\hat{\mu}_{j,d,a}$ and standard deviation $\hat{\sigma}_{j,d,a}$. The joint probability is computed by treating each component as conditionally independent given an image $I_j$:

$$\tilde{Pr}(h_1, \ldots, h_D \mid G_{I_j} = g) = \prod_{d=1}^{48} \mathcal{N}(h_d \mid \hat{\mu}_{j,d,a}, \hat{\sigma}_{j,d,a}) \quad (4.16)$$

So far we have presented overall meta-learning framework. In following sections we present the experimental set-up and results.

## 4.3 Experimental Setup

### 4.3.1 Dataset

Training of meta-learner requires sufficient number of images for which ground-truth is available. For this reason we have used Wang dataset for our experiments. This dataset consist of 10 categories and each category consist of 100 images. Images in each category are semantically similar to remaining 99 images in that category. It means that for each one of the 1000 images, we have 99 ground-truth images. Since this dataset has large number of images for which ground-truth is available, we have enough number of images and their corresponding ground-truth to train our meta-learner. We have used 600 of these images and their corresponding ground-truths to train the meta-learner. Remaining 400 images are used to test the performance of meta-learner and compare it against the performance of black-box.

### 4.3.2 Image Features

We have used salient point based local feature vectors. Like previous chapter , we have extracted 23-dimensional feature vector from fixed size blocks surrounding each salient point. The type of features extracted are same as described in section 3.4.2.

### 4.3.3 Black-box

Our notion of black-box comprise of two things:

- Matching mechanism : to find the ordered set of images that are semantically similar to any given query image.

| Mean Average Precision (MAP) values | | |
| --- | --- | --- |
| Black-box name | Performance of Black-box | Performance of Meta-learner |
| Gaussian+C2(Top 100) | .2425 | **.5590** |
| 2GMM+C2(Top 100) | .2045 | **.5419** |
| IMK(Top 100) | .2138 | **.5487** |

Table 4.1: Comparison of performances of black-box and meta-learner for image retrieval task

- Threshold : number of top *T* images from the above mentioned ordered set of images that are presented to the user as a retrieval result.

e.g. if the method for finding semantically similar image to a query image is IMK based matching and top 100 image from the ordered output of IMK based matching system are rendered to user, then we call this black-box as *IMK(Top 100)*.

In our experiments we have used following three matching mechanism to find ordered set of semantically similar images to any given query image:

**Gaussian+C2** Each image is modeled using Gaussian pdf (with diagonal covariance) and C2 divergence is used to compute similarity between pdf's of two images.

**2GMM+C2** Each image is modeled using 2 component Gaussian mixture model and C2 divergence is used to compute similarity between pdf's of two images.

**IMK** IMK is used to compute similarity between two images.

## 4.4 Result and Discussions

Table 4.1 shows the performance of different black-boxes and corresponding meta-learners (build on top of them) in terms of Mean average precision values computed over 400 test query images. We observe that use of meta-learner increases the image retrieval performance more than two times across all underlying black-boxes. Figure 4.2 shows some query images, retrieval result of black-box (IMK(Top 100)) and corresponding meta-learner. The improvement in the retrieval results is noticeable.

Now, for every test query image individually, we analyze the difference between performance of black-boxes and corresponding meta-learners. Let average precision value obtained for a test query image using meta-learner be $ap_{meta}$ and using black-box be $ap_{bbox}$. Figure 4.2 shows difference between $ap_{meta}$ and $ap_{bbox}$ versus different test query images for all three black-boxes and their corresponding meta-learners mentioned in Table 4.1. We can see that for some of the test query images difference between $ap_{meta} - ap_{bbox}$ is less than zero, i.e. for some test query images meta-learner is performing inferior to underlying black-box. Table 4.3 present the number of test query images (out of total 400 test query images) for which performance of meta-learner is inferior to black-box.

Let $N(X)$ represent number of test query images for which $(ap_{meta} - ap_{bbox}) \geq X$. Figure 4.3 represent the relationshiop between $N(X)$ and $X$ for different black-boxes. We observe that the value of $N(X)$ is lower for black-boxes using more sophisticated matching mechanism (i.e. for black-boxes 2GMM+C2(Top100) and IMK(Top100)) than for black-box using simple matching mechanism (i.e for black-box Gaussian+C2(Top100)). It may be because sophisticated black-boxes are better capable of modeling semantic similarity between images and in turn meta-learner build on top of them are more stable in terms of performance across different query images.

Now, we present the effect on performance of meta-learner upon change in *Threshold* value (as mentioned in section 4.3.3) of black-box. Figure 4.4 shows the performance of meta-learners and underlying black-boxes for various values of threshold $T$. We observe that for performance of all the three black-boxes increases monotonically with increase in threshold value, which is expected. Whereas, performance of meta-learners degrades if threshold values approaches the middle of threshold range(Note : Threshold range for any black-box is 0 to *Number of images in search space*). Ideally, it should also increase with increase in threshold value. The reason behind this behavior may be that when the threshold value approaches the middle of threshold range, the difference between number of zeros and ones in function $f_{bbox}(I) = (R_{I_1}, .........., R_{I_K})$ decreases and because of this deceasing difference the knowledge acquired by meta-learner about underlying black-box is decreased, which in turn results in degraded performance of meta-learner. Also, let the consistency in performance of the meta-learner (across all values of threshold) build on the black-box $bbox(TopT)$, be denoted by $C(bbox(TopT))$, then according to Figure 4.4:

$$C(IMK(TopT)) > C(2GMM + C2(TopT)) > C(Gaussian + C2(TopT)) \qquad (4.17)$$

| Test query image | Top 5 images retrieved by IMK(Top 100) | | | | | Top 5 images retrieved by Meta-Learner | | | | |
|---|---|---|---|---|---|---|---|---|---|---|
| 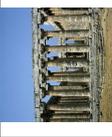 | 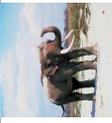 | 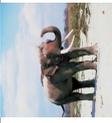 | 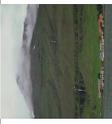 | 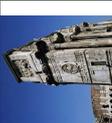 | 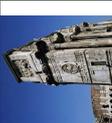 | 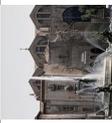 | 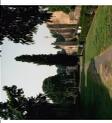 | 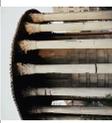 | 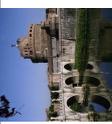 | 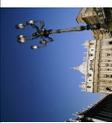 |
| 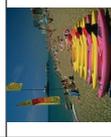 | 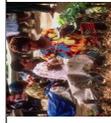 | 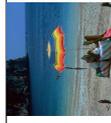 | 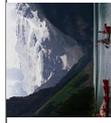 | 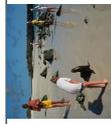 | 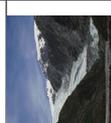 | 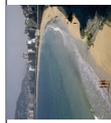 | 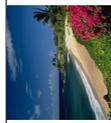 | 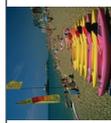 | 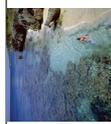 | 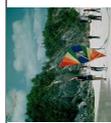 |
| 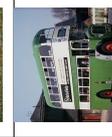 | 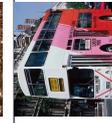 | 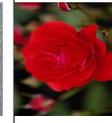 | 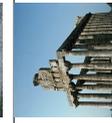 | 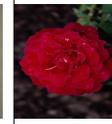 | 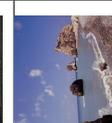 | 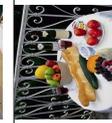 | 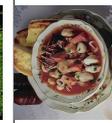 | 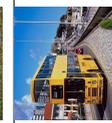 | 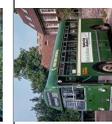 | 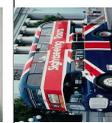 |
| 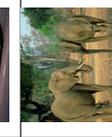 | 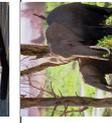 | 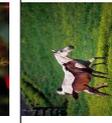 | 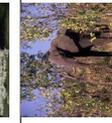 | 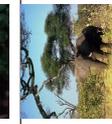 | 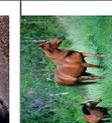 | 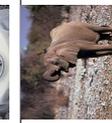 | 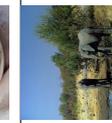 | 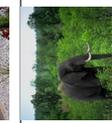 | 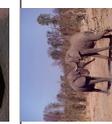 | 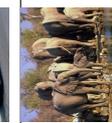 |
| 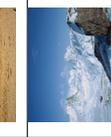 | 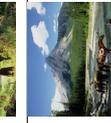 | 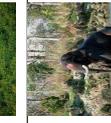 | 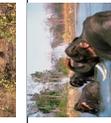 | 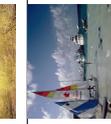 | 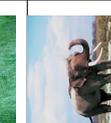 | 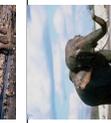 | 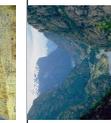 | 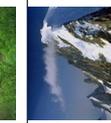 | 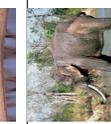 | 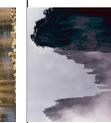 |

Table 4.2: Example test query images and top 5 retrieval results of IMK(Top 100) and corresponding meta-learner

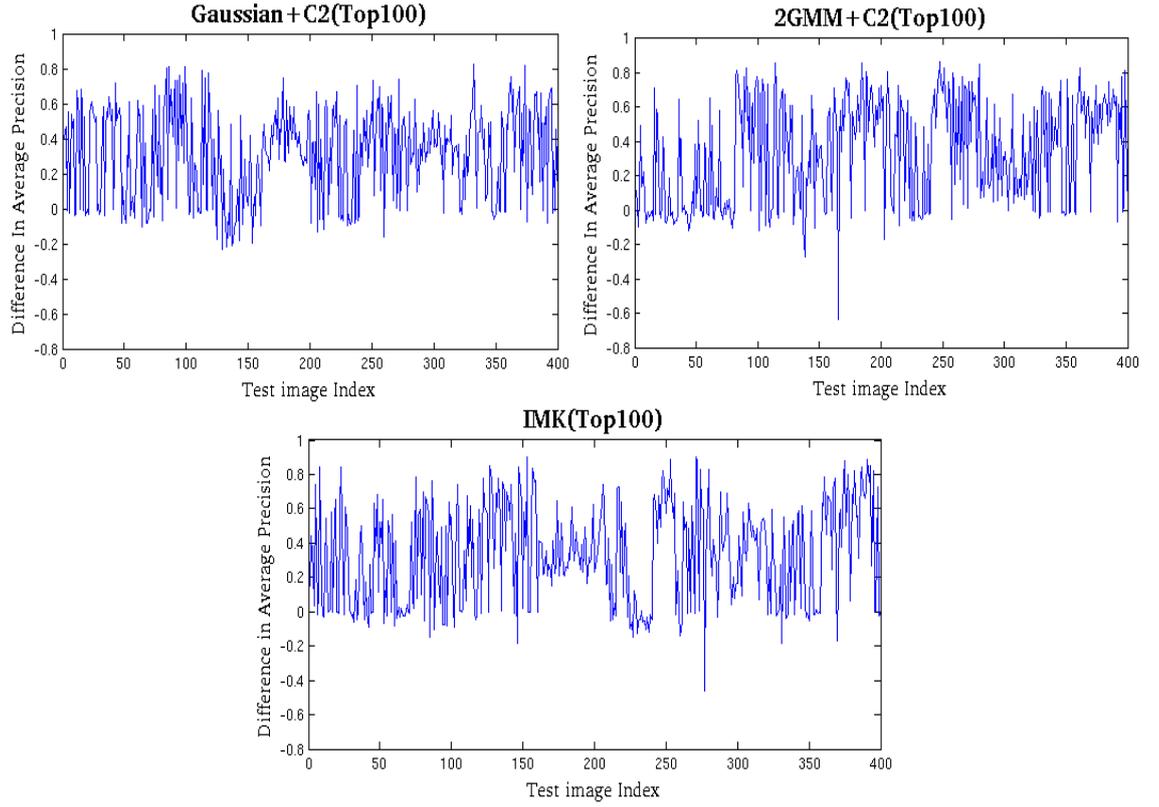

Figure 4.2: Difference in Average precision obtained using meta-learner and black-box

| Black-box name | Number of test query images for which meta-learner performs inferior to black-box |
|---|---|
| Gaussian+C2(Top 100) | 91 |
| 2GMM+C2(Top 100) | 98 |
| IMK(Top 100) | 90 |

Table 4.3: Number of test query images for which meta-learner performs inferior to black-box

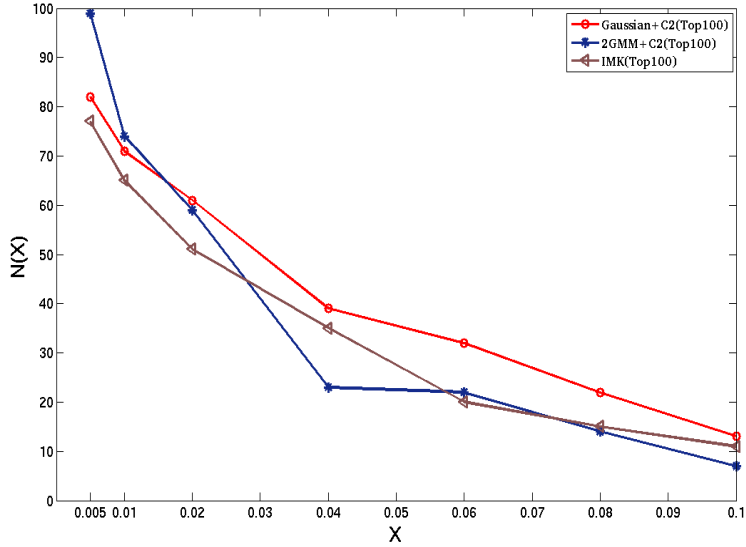

Figure 4.3: *N(X) versus X*

This reinforces our belief that performance of meta-learner is more consistent for more sophisticated matching mechanism.

## 4.5 Summary

We presented a meta-learning framework for image retrieval. This framework is independent of the underlying black-box. We analyzed the performance of meta-learners build on three different black-boxes. Experiments show that meta-learners build on top of any of these black-boxes results in substantial performance improvement.

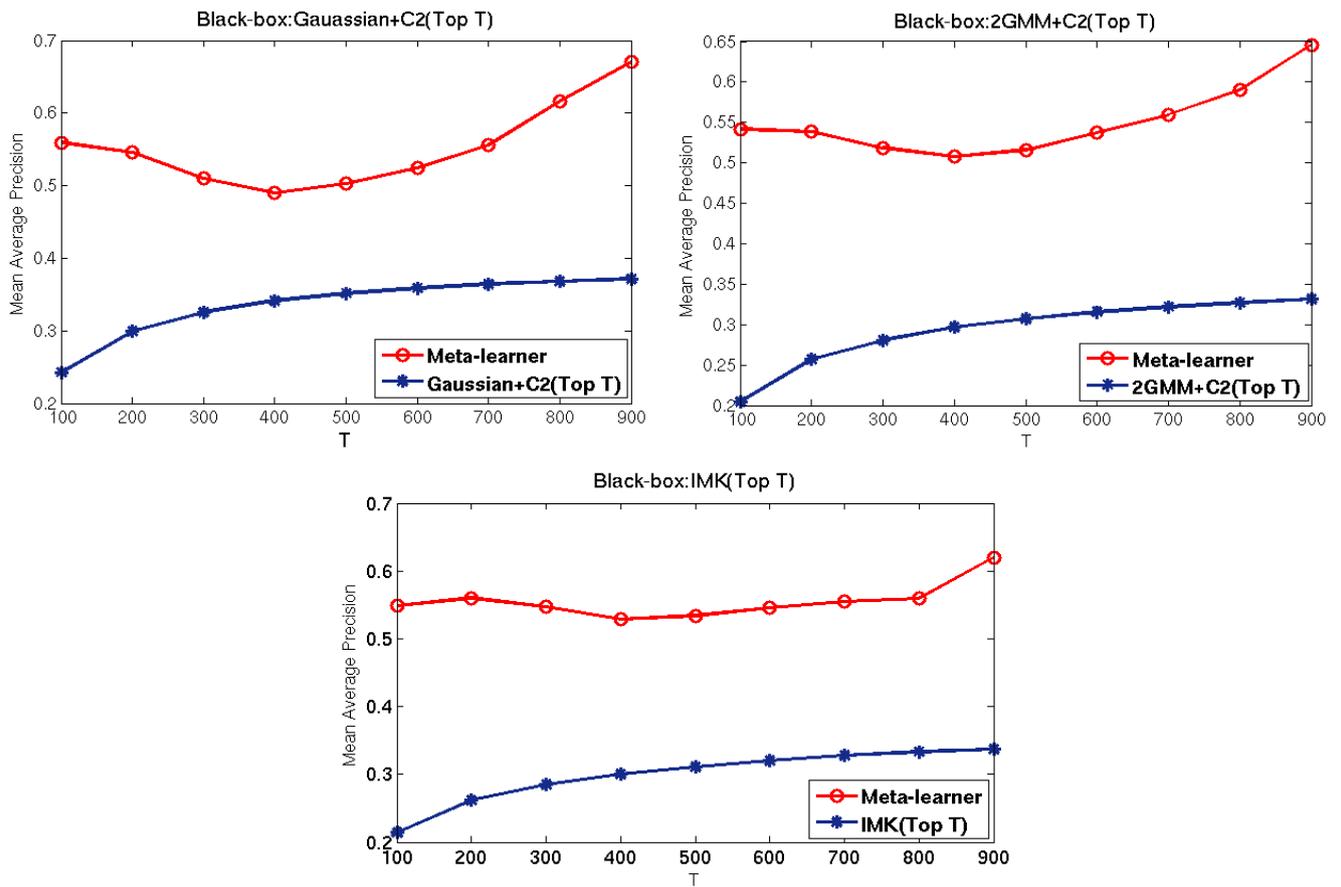

Figure 4.4: Effect of changing Threshold value on performance of meta-learner

# Chapter 5

# IMAGE CLASSIFICATION USING VARIABLE LENGTH PATTERNS

In this chapter we shift our focus on image classification using variable length patterns. When images are represented using variable length patterns, the natural choice of classification model is generative models (like GMM or Conditional random field(CRF) etc). Other models which use fixed length patterns (like SVM ,kNN) cannot be readily used for varying length patterns. They either require varying length patterns to be converted to fixed length patterns using techniques such as bag-of-word(BoW) or they require use of some function that takes as input two varying length patterns and computes similarity (or dissimilarity) between them (like dynamic kernels) . After this, any discriminative classifier can be applied to this representation for classification.

BoW representation based classifiers have shown promising performance and have been in vogue for quite some time now (48; 49; 50). As shown in Bosch et al. (51), better results than BoW representation based classifier can be obtained if first images are represented in intermediate semantic space and then any discriminative classifier is used. Bosch et al. (51) uses probabilistic Latent Semantic Analysis (pLSA) to obtain a intermediate semantic space representation of images. The limitation with pLSA model is that the number of parameters grow linearly with the size of the dataset, which leads to serious problem of over-fitting. To overcome this limitation, we propose to use Latent Dirichlet Allocation to obtain intermediate semantic representation of images and subsequently apply any discriminative classifier on this representation for classification of images.

Despite the promising performance of BoW based classifier, they have a limitation that stems from the fact that, BoW representation of images is based on quantization of local feature vectors, which inherently causes classification model build upon this representation to suffer from some loss of information. Moreover, since LDA representation of images is constructed using BoW representation as input, classification models build upon this representation also suffer from loss of information. In contrast to this, the classification models which use dynamic kernels (like Intermediate Matching kernel) as a measure of similarity do not incur any loss in information due to quantization.

## 5.1 Objective of this Chapter

In this chapter we have investigated answers to following questions:

- Whether use of LDA representation in discriminative classifiers gives better performance than use of BoW representation in discriminative classifiers ?

- How well the discriminative classifiers that use dynamic kernel ( which incur no information loss) perform against discriminative classifier that use BoW or LDA representation (which incur information loss due to quantization) ?

Figure 5.1 represents the framework for above mentioned three classification approaches.

Rest of the chapter is organized as follows: Section 5.2 and 5.3 explain method for obtaining BoW and LDA representation of images respectively. Section 5.4 and 5.5 explains kNN and SVM classifier for multiclass classification respectively. Section 5.6 and Section 5.7 present the experimental setup and results respectively.

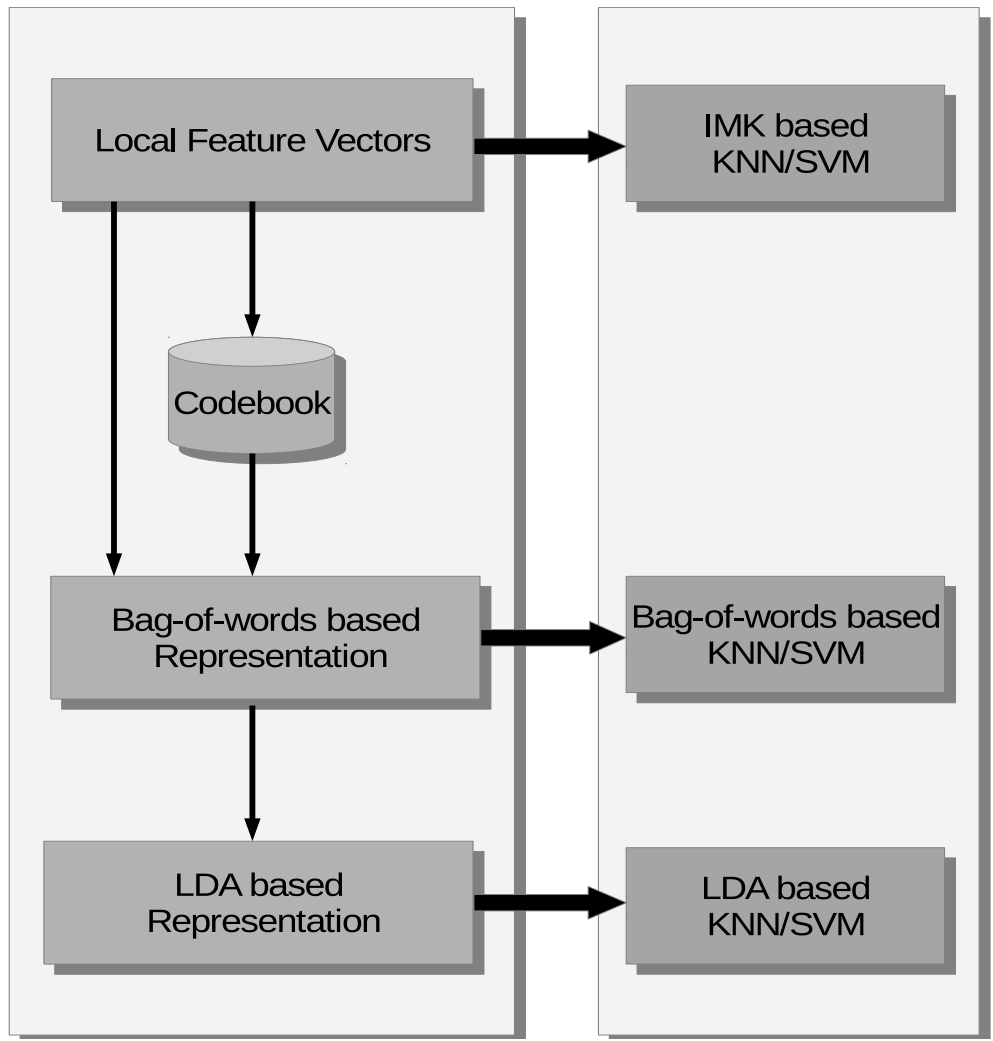

Figure 5.1: Framework for classification of images with varying length patterns

## 5.2 BoW Representation of Images

For creating BoW representation of images, first a codebook is constructed by applying a clustering algorithm on local features. Generally, k-means algorithm is used for clustering. The cluster centers are considered as "visual words" in codebook. Each feature in an image is then quantized to closest visual word in codebook and entire image is represented as global histogram of number of occurrences of each visual word in the codebook. The size of the resulting histogram is equal to number of words in the codebook and hence number of clusters obtained from clustering algorithm. Mathematically, let $c_1, .......c_n \in \mathbb{R}^d$ be the $n$ visual words (cluster centers) and let $\mathbf{x}_1, .......\mathbf{x}_m \in \mathbb{R}^d$ be $m$ local feature extracted from an image (where $d$ is the dimensionality of local feature vector). The BoW representation of image is $n$-dimensional vector $\mathbf{w}$ and the $i_{th}$ component $w_i$ of $\mathbf{w}$ is computed as:

$$w_i = \sum_{j=1}^{m} \alpha(i, \arg\min_{k \in \{1....n\}} \|x_j - c_k\|) \tag{5.1}$$

where $\alpha(a, b) = 1$ if $a = b$ and 0 otherwise, $\|.\|$ is the euclidean distance. Note that $\sum_{i=1}^{n} w_i = m$, since each feature vector is assigned to only one visual codeword.

## 5.3 LDA Representation of Images

In this section we briefly present LDA model (53) along with inferencing technique and performance measure for this model.

### 5.3.1 Latent Dirichlet Allocation

Latent Dirichlet Allocation is an unsupervised generative probabilistic model that discovers latent semantic topics in the dataset. It is a three-level hierarchical Bayesian model, in which each document (images in our case) of dataset is modeled as a finite mixture over an underlying set of topics. Each topic is, in turn, is characterized by its own particular distribution over words (visual words in our case). The intuition behind is that words bear strong semantic information and items discussing similar topics will use a similar group of words. Latent topics are thus discovered by identifying groups of words in the dataset that frequently occur together within documents.

For explaining the process of modeling a dataset of images using LDA, we use the following terminology and notation:

A *visual word* $w \in \{1, ........, V\}$ is the most basic unit of dataset. For cleaner notation, we represent words using unit-basis vectors that have a single component equal to 1 and all other components equal to 0. Thus, using subscript to denote component, the $v^{th}$ word in the vocabulary is represented by a $V$-vector $w$ such that $w^v = 1$ and $w^u = 0$ for $u \neq v$.

An *image* is a sequence of $N$ visual words denoted by $\mathbf{w} = (w_1, w_2, ......., w_N)$, where $w_n$ is the $n^{th}$ visual word in sequence.

A *dataset* is a collection of $M$ images denoted by $\mathcal{D} = \{\mathbf{w}_1, \mathbf{w}_2, ......, \mathbf{w}_M\}$.

A *topic* $z \in \{1, ....., K\}$ is a probability distribution over a vocabulary of $V$ visual words.

Now we are ready to define the LDA model. In LDA visual words are modeled as observed random variables and topics are modeled as latent (hidden) random variable. First we describe the generative process assumed by LDA for each image $\mathbf{w}$ in a dataset $\mathcal{D}$. Once the generative procedure is established, we may define its joint distribution and then use statistical inference to compute the probability distribution over the latent variables, conditioned on the observed variables. The generative process for each image $\mathbf{w}_m, m \in \{1, .....M\}$ in a dataset $\mathcal{D}$ is as follows:

1. Choose a K-dimensional topic weight vector $\theta_m$ from the distribution $p(\theta|\alpha) = Dirichlet(\alpha)$.

2. For each visual word $w_n$, $n \in \{1,......., N\}$ in an image:
   (a) Choose a topic $z_n \in \{1,......., K\}$ from the multinomial distribution $p(z_n = k|\theta_m) = \theta_m^k$.
   (b) For the chosen topic $z_n$, draw a word $w_n$ from the probability distribution $p(w_n = i|z_n = j, \beta)$

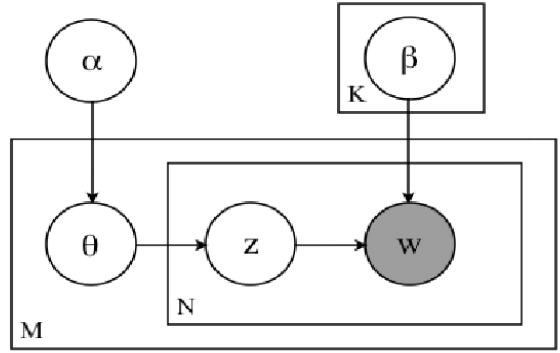

Figure 5.2: Graphical representation of LDA

We assume that the dimensionality $k$ of the Dirichlet distribution (and thus the dimensionality of the topic variable $z$) is known and fixed. $\beta$ is a $K \times V$ matrix that parameterizes the word probabilities, where $\beta_{ij} = p(w^j = 1|z^i = 1)$. This matrix is to be estimated using dataset.

A $k$-dimensional Dirichlet random variable $\theta$ can take values in the $(k-1)$-simplex (a $k$-vector $\theta$ lies in the $(k-1)$-simplex if $\theta_i \geq 0, \sum_{i=1}^{k} \theta_i = 1$), and has the following probability density on this simplex:

$$p(\theta|\alpha) = \frac{\Gamma(\sum_{i=1}^{K} \alpha_i)}{\prod_{i=1}^{K} \Gamma(\alpha_i)} \theta_1^{\alpha_1 - 1} ....... \theta_K^{\alpha_k - 1} \tag{5.2}$$

where the parameter $\alpha$ is a $k$-vector with components $\alpha_i > 0$, and where $\Gamma(x)$ is the Gamma function.

Assuming for now that parameters $\alpha$ and $\beta$ are given, the joint distribution over the topic mixtures $\theta$ and the set of $N$ topics $\mathbf{z}$ and a set of $N$ words $\mathbf{w}$ is given by:

$$p(\theta, \mathbf{z}, \mathbf{w}|\alpha, \beta) = p(\theta|\alpha) \prod_{n=1}^{N} p(z_n|\theta) p(w_n|z_n, \beta) \tag{5.3}$$

Figure 5.2 represents the Graphical representation of LDA. Note that the parameter $\alpha$ is a $K$-dimensional parameter that is same for all of the images within a dataset. Since we extract our topic mixture weights from $\alpha$, an extra level of indirection is introduced that allows us to analyze a previously unseen image in relation to the rest of the existing images in the corpus. This overcomes the limitations of the pLSI model, which treats the topic mixture weights as a large set of individual parameters that are explicitly linked to images in the training set.

## 5.3.2 Inferencing

Parameters $\alpha$ and $\beta$ completely specify the model given in Equation. 5.3. Now, the key problem to be solved for making use of LDA for dataset modeling is that of computing posterior distribution of latent topic variables given a document. Bayes rule is applied for this purpose as follows:

$$p(\theta, \mathbf{z}|\mathbf{w}, \alpha, \beta) = \frac{p(\theta, \mathbf{z}, \mathbf{w}|\alpha, \beta)}{p(\mathbf{w}|\alpha, \beta)} \tag{5.4}$$

The denominator of above equation is likelihood of an image $p(\mathbf{w}|\alpha, \beta)$ and it is obtained by integrating Equation 5.3 over $\theta$ and summing over $z$ as follows:

$$p(\mathbf{w}|\alpha, \beta) = \int p(\theta|\alpha) \left\{ \prod_{n=1}^{N} \sum_{z_n} p(z_n|\theta) p(w_n|z_n, \beta) \right\} d\theta \tag{5.5}$$

Marginalization over the latent variables and writing Equation 5.5 gives:

$$p(\mathbf{w}|\boldsymbol{\alpha},\boldsymbol{\beta}) = \frac{\Gamma(\sum_i \alpha_i)}{\prod_i \Gamma(\alpha_i)} \int \left\{ \prod_{i=1}^{K} \theta_i^{\alpha_i - 1} \right\} \left\{ \prod_{n=1}^{N} \sum_{i=1}^{K} \prod_{j=1}^{V} (\theta_i \beta_{ij})^{w_n^j} \right\} d\boldsymbol{\theta} \qquad (5.6)$$

Above function is intractable due to the coupling between $\theta$ and $\beta$ in summation over latent topics and hence the posterior distribution is intractable.

To overcome this problem, approximate probabilistic inference is used. This approximation consists of two steps. First, a tractable family of distributions is chosen, whose statistics are easy to compute. In next step, we attempt to select the particular member of this family that best approximates the true posterior distribution. This is done by computing the optimal variational parameters, $\gamma^*$ and $\phi^*$, that minimizes the difference between the variational distribution and true posterior distribution. The optimizing parameters $(\gamma^*(\mathbf{w}), \phi^*(\mathbf{w}))$ are document-specific. It means, the Dirichlet parameters $\gamma^*(\mathbf{w})$ are used to provide a representation of an image in the topic simplex.

### 5.3.3 Measure for Evaluating the Document Modeling Performance of LDA

The simplest way to evaluate the performance of a document modeling framework like LDA is to compute its perplexity on a held-out test set. Perplexity is defined as:

$$perplexity(\mathcal{D}_{test}) = exp\left\{ -\frac{\sum_{m=1}^{M} \log p(d_m)}{\sum_{m=1}^{M} N_m} \right\} \qquad (5.7)$$

where $N_m$ is the number of words in the $m_{th}$ document of corpus $\mathcal{D}_{test}$. Lower values of perplexity indicates better generalization performance of model.

## 5.4 k-Nearest Neighbor(kNN) Classifier

k-Nearest Neighbor classifier does not require any model to be fitted. This classifier is based on *learning by analogy*, that is, given a query data $\mathbf{x}_0$, we find $k$ training data point $\mathbf{x}_{(i)}, i = 1, \ldots .k$ closest to query data $\mathbf{x}_0$, and then classify using majority vote among the $k$ neighbors. Ties are broken at random. "Closeness" between two data points can be defined in term of any distance metric, like euclidean distance, mahalanobis distance or kernel function between two data points. Typically, we normalize each of the feature to have mean zero and variance 1.

Despite its simplicity kNN has been shown to successful in large number of classification problems including satellite image scene, handwritten digits etc.

## 5.5 Support Vector Machine(SVM)

The SVM (52) is a linear two-class classifier. The principle of SVM is based on **Cover's theorem**. The Cover's theorem states that, a complex pattern classification problem cast non-linearly into a high-dimensional space is more likely to be linearly separable than in a low-dimensional space. Figure 5.3 shows example of a pattern classification problem that is non-linearly separable in a 2-dimensional input space but linearly separable in a 3-dimensional feature space

The non-linearly separable data in an input space $\mathbb{X}$ can be mapped to a higher-dimensional space $\mathbb{Z}$ using a non-linear function $\phi(\mathbf{x})$. This higher dimensional space, $\mathbb{Z}$, is called a *feature space*. The aim of SVM is to construct an *optimal hyperplane* that separates the examples of two classes. The problem of constructing the optimal hyperplane is posed as the problem of finding the maximum margin hyperplane. The margin is the distance of a hyperplane to the nearest example and is given as $\frac{1}{\|\mathbf{w}\|}$, where $\mathbf{w}$ is the weight vector of

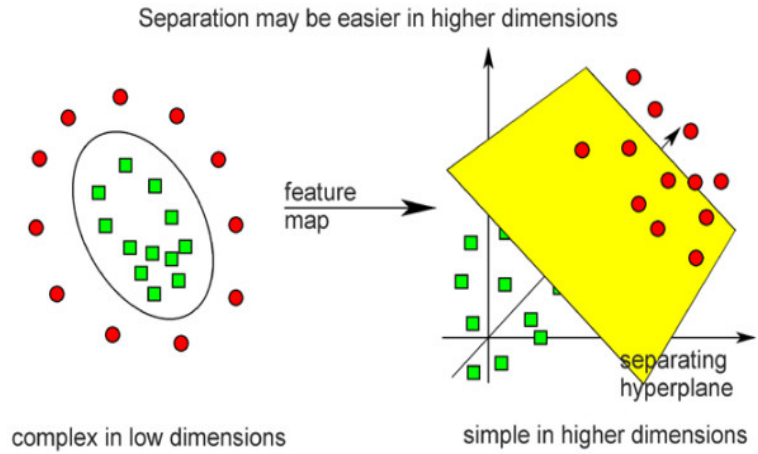

Figure 5.3: Example of a pattern classification problem that is non-linearly separable in a 2-dimensional input space but linearly separable in a 3-dimensional feature space

the hyperplane. Given a dataset consisting of $n$ data points, $(\mathbf{x}_i, y_i), i = 1, 2, \ldots, n$, where $\mathbf{x}_i$ is the $i^{th}$ data point, $y_i \in (+1, -1)$ denotes the class label of $i^{th}$ data point. A hyperplane is defined by:

$$\mathbf{w}^T \phi(\mathbf{x}) + b = 0 \tag{5.8}$$

where b is the bias. To construct an optimal hyperplane, $\|\mathbf{w}\|$ has to be minimized. The problem of constructing the optimal hyperplane is formulated as:

$$\text{minimize } J(\mathbf{w}, \xi) = \frac{1}{2}\|\mathbf{w}\|^2 + C \sum_{i=1}^{n} \xi_i \tag{5.9}$$

such that $y_i(\mathbf{w}^T \mathbf{z}_i + b) \geq 1 - \xi_i, i = 1, 2, \ldots, n$

and $\xi_i \geq 0, i = 1, 2, \ldots n$

Here $\mathbf{z}_i$ is the feature space vector $\phi(\mathbf{x}_i)$ for $x_i$, $\xi_i$ are the slack variables and $C$ is a positive trade-off parameters specified by the user. Solving the above problem by quadratic optimization, the discrimination function for the optimal hyperplane is defined

as:

$$D(\mathbf{x}) = \mathbf{w}^T \phi(\mathbf{x}) + b = \mathbf{w}^T \mathbf{z} + b = \sum_{i=1}^{n_s} \lambda_i y_i \mathbf{z}^T \mathbf{z}_i + b \qquad (5.10)$$

where $\lambda_i$ are the Lagrangian multiplier coefficients and $n_s$ is the number of *Support Vectors*. The support vector define the optimal separating hyperplane. The non-linear transformation of a vector in the input space to a vector in the feature space can be achieved using a inner-product *kernel* function, $K(.,.)$, defined as:

$$K(\mathbf{x}_i, \mathbf{x}_j) = \phi(x_i)^T \phi(\mathbf{x}_j) = \mathbf{z}_i^T \mathbf{z}_j \qquad (5.11)$$

The kernel function should satisfy *Mercer's theorem*. The architecture of SVM is shown in Figure 5.4.

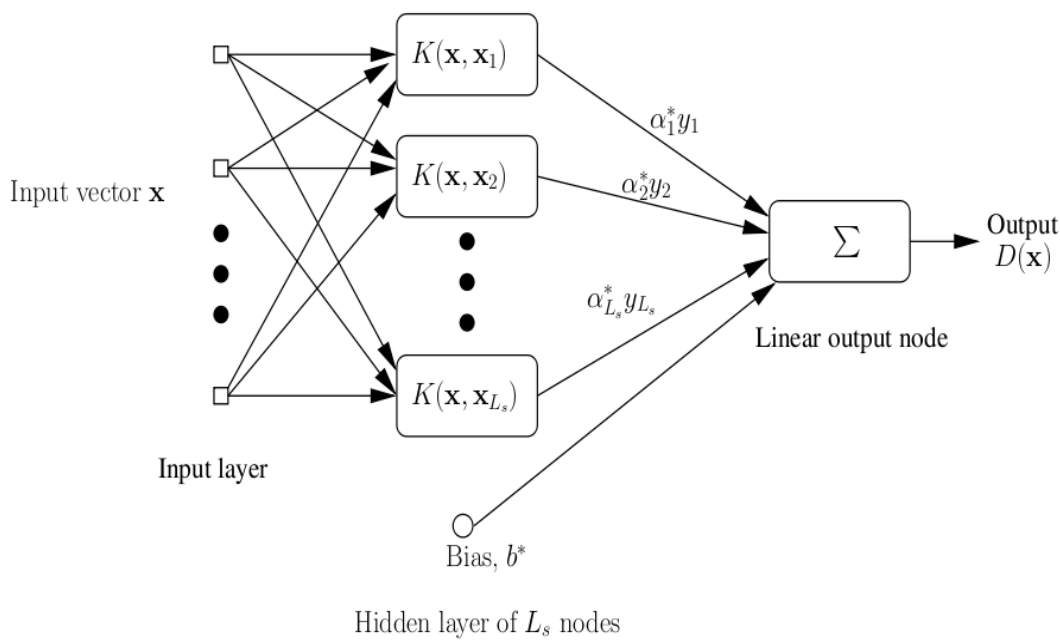

Figure 5.4: Architecture of a support vector machine for two-class pattern classification. The class of the input pattern **x** is given by the sign of the discriminant function $D(\mathbf{x})$. The number of hidden nodes corresponds to the number of support vectors $L_s$. Each hidden node computes the inner-product kernel function $K(\mathbf{x}, \mathbf{x}_i)$ on the input pattern **x** and a support vector $\mathbf{x}_i$.

## 5.5.1 Multiclass Pattern Classification using SVMs

Support vector machines are originally designed for two-class pattern classification. Multiclass pattern classification problems are commonly solved using a combination of two-class SVMs and a decision strategy to decide the class of the input pattern (54). Now we present the two approaches to decomposition of the learning problem in multiclass pattern classification into several two-class learning problems so that a combination of SVMs can be used. The training data set $\{(\mathbf{x}_i, c_i)\}$ consists of $L$ examples belonging to $T$ classes. The class label $c_i \in \{1, 2, ..., T\}$. For the sake of simplicity, we assume that the number of examples for each class is the same, *i.e.*, $L/T$.

**One-against-the-rest approach**

In this approach, an SVM is constructed for each class by discriminating that class against the remaining ($T$-1) classes. The classification system based on this approach consists of $T$ SVMs. All the $L$ training examples are used in constructing an SVM for each class. In constructing the SVM for the class $t$ the desired output $y_i$ for a training example $\mathbf{x}_i$ is specified as follows:

$$\begin{aligned} y_i &= +1, & \text{if} \quad c_i = t \\ &= -1, & \text{if} \quad c_i \neq t \end{aligned} \quad (5.12)$$

The examples with the desired output $y_i = +1$ are called *positive* examples. The examples with the desired output $y_i = -1$ are called *negative* examples. An optimal hyperplane is constructed to separate $L/T$ positive examples from $L(T\text{-}1)/T$ negative examples. The much larger number of negative examples leads to an imbalance, resulting in the dominance of negative examples in determining the decision boundary. The extent of imbalance increases with the number of classes and is significantly high when the number of classes is large. A test pattern $\mathbf{x}$ is classified by using the *winner-takes-all* strategy that uses the

following decision rule:

$$\text{Class label for } \mathbf{x} = \arg\max_t D_t(\mathbf{x}) \qquad (5.13)$$

where $D_t(\mathbf{x})$ is the discriminant function of the SVM constructed for the class $t$.

**One-against-one approach**

In this approach, an SVM is constructed for every pair of classes by training it to discriminate the two classes. The number of SVMs used in this approach is $T(T\text{-}1)/2$. An SVM for a pair of classes $s$ and $t$ is constructed using $2L/T$ training examples belonging to the two classes only. The desired output $y_i$ for a training example $\mathbf{x}_i$ is specified as follows:

$$\begin{aligned} y_i &= +1, \quad \text{if} \quad c_i = s \\ &= -1, \quad \text{if} \quad c_i = t \end{aligned} \qquad (5.14)$$

The small size of the set of training examples and the balance between the number of positive and negative examples lead to a simple optimization problem to be solved in constructing an SVM for a pair of classes. When the number of classes is large, the proliferation of SVMs leads to a complex classification system.

The *max-wins* strategy is commonly used to determine the class of a test pattern $\mathbf{x}$ in this approach. In this strategy, a majority voting scheme is used. If $D_{st}(\mathbf{x})$, the value of the discriminant function of the SVM for the pair of classes $s$ and $t$, is positive, then the class $s$ wins a vote. Otherwise, the class $t$ wins a vote. Outputs of SVMs are used to determine the number of votes won by each class. The class with the maximum number of votes is assigned to the test pattern.

## 5.6 Experimental Setup

### 5.6.1 Tools used

We have used LIBSVM tool for SVM classification. For codebook creation, we have used mpi-k-means-1.5's implementation of k-means algorithm.

### 5.6.2 Dataset

We have selected 5 image categories from Large Scale Visual Recognition Challenge 2010 (ILSVRC2010) namely Seashore (2382 training images), Alp (2005 training images), Bridge (1598 training images), Castle (1379 training images), Skyscraper (1546 training images). For each of these categories 50 validation images and 150 test images are available. We have chosen 400 images randomly from each category for training. The images are real-world and with high intra-class variability.

### 5.6.3 Image Features

Along with images, ILSVRC also provides local feature vectors for each image. Prior to feature extraction, each image is resized to have a max side length of no more than 300 pixel, then SIFT descriptors (32) are computed on 20x20 overlapping patches with a spacing of 10 pixels. Images are further downsized (to 1/2 the side length and then 1/4 of the side length) and more descriptors are computed. On average, each image consists of 700 local feature vectors.

### 5.6.4 Details of Classification Systems

Details of various classification systems are as follows:

- Both in case of BoW based kNN and LDA based kNN, we have used euclidean distance as a measure of similarity.

- In case of IMK based kNN, we have taken IMK between two images as a measure of similarity.

- Both in case of BoW based SVM and LDA based SVM, Radial Basis Function (RBF) kernel is used as kernel function between two images.

- In case of IMK based SVM, we have chosen 128 component and 256 component UBM for construction of set of virtual feature vectors. RBF is chosen as base kernel for computing kernel function between selected feature vectors.

- All SVM classifiers are trained using one-against-one approach, since number of categories is 5 and only 10 binary SVM's are needed to be trained for one multiclass SVM. Also, as mentioned in Section 5.5.1 it has also the advantage that number of positive and negative examples in each binary SVM is same.

## 5.7  Results and Discussion

We have set the codebook size for BoW representation to be 1000. Thus the dimensionality of BoW representation is 1000. In context of LDA representation, we first need to select number of latent topics in LDA model. For this we have trained LDA model with different number of latent topics and calculated the perplexity of LDA model on previously unseen 500 images (100 images from each category). Figure 5.5 shows the graph between perplexity computed on previously unseen images and number of latent topics in LDA.

| Classification system name | Classification Accuracy (%) |
|---|---|
| BoW based kNN | 45.86 |
| LDA based kNN | 57.06 |
| IMK based kNN with number of components in UBM=128 | 58.66 |
| IMK based kNN with number of components in UBM=256 | 60.13 |

Table 5.1: Classification accuracy of kNN classifier using different image representations

We observe that value of perplexity is minimum for number of topic equal to 120. So we have have used 120 latent topics for obtaining the LDA representation of images. As described in Section 5.3.2 the LDA representation of an image will be a 120-dimensional vector of dirichlet parameters $\gamma^*(\mathbf{w})$.

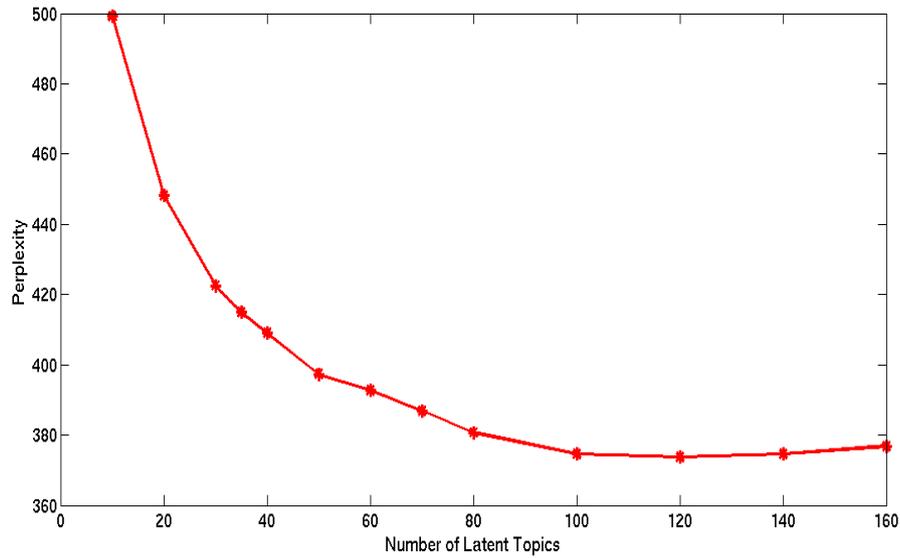

Figure 5.5: Perplexity vs Number of latent topics in LDA

Table 5.1 shows performance of kNN classifier using different image representation.

Table 5.2 shows performance of SVM classifier using different image representation.

Results show that use of LDA representation instead of BoW representation produces much better (more than 11%) accuracy in case of kNN classifier. In case of SVM classifiers, again, the use of LDA representation results in better accuracy than use of BoW representation. But the difference between LDA based SVM and BoW based SVM is not as much as difference between LDA based kNN and BoW based kNN.

| Classification system name | Classification Accuracy (%) |
|---|---|
| BoW based SVM | 71.86 |
| LDA based SVM | 74.06 |
| IMK based SVM with number of components in UBM=128 | 75.33 |
| IMK based SVM with number of components in UBM=256 | 78.53 |

Table 5.2: Classification accuracy of SVM classifier using different image representation

IMK based kNN works better than both BoW based kNN and LDA based kNN. Also, IMK based SVM works better than both BoW based SVM and LDA based SVM. This may be because unlike IMK, both BoW and LDA representation are based on quantized features and thus they incur information loss.

## 5.8 Summary

In this chapter, we have presented different techniques for image classification when images are represented as variable length patterns. We have shown that use of LDA representation instead of BoW representation in discriminative classifier significantly increases the performance of classifier. Also IMK based classifier are better than both BoW and LDA based classifiers.

# Chapter 6

# CONCLUSION

In Chapter 3 of this work, One Step Matching and Two Step Matching approaches using Intermediate Matching Kernel for the task of image retrieval have been proposed. Experiments have shown that use of UBM based IMK for similarity computation gives much better MAP than various baseline systems. Use of center of cluster based IMK is not as good as UBM based IMK for image retrieval (almost 25% degradation in performance). In Two Step Matching , we have proposed to use more than one closest cluster as potential search space. We have also specified the motivation for this proposal in Section 3.1.2. Using more than one closest cluster as potential search space is shown to have almost **50**% better retrieval performance than using only one closest cluster as potential search space. This justifies our intuition in Section 3.1.2.

Chapter 4 of this work explains the development of a Meta-learning framework for improving the retrieval performance of an existing image retrieval system. This Meta-learning framework makes use of available ground-truth, output of underlying image retrieval system. Experiments have shown that this Meta-learning framework gives almost twice as good as performance than underlying image retrieval system.

Chapter 5 deals with image classification using varying length patterns. In this chapter, we explain methods for obtaining BoW based representation and LDA based representation of images. Once these representations are obtained, these are fed to the classifiers like kNN and SVM. This gives us BoW based and LDA based classification systems. At the end of this chapter, the classification accuracies of these system are compared with accuracy obtained using IMK based classifier. Experiments show that IMK based classifier outperform BoW-based and LDA-based classifiers.

## 6.1 Scope for Future Work

In the context of IMK computation, if labeled data is available, we can compute class-specific UBM's instead of UBM computed from whole data. Use of class-specific UBM based IMK is supposed to give better performance than use of whole data UBM based IMK since we are using extra information of image labels in UBM computation.

In our Meta-learning framework for image retrieval, the function $f_{bbox}$ maps all the images retrieved by underlying black-box to 1 and images not retrieved by underlying black-box to 0. Thus, this function does not make use of the ranking order of images retrieved by underlying black-box. In future work, Meta-learning framework can be modified such that it makes use of ranking order information in function $f_{bbox}$. Use of ranking order should increase the performance of Meta-learning framework.

LDA model is unable to model correlation between latent topics and assumes that latent topics are independent to each other. This limitation stems from the use of the Dirichlet distribution to model the variability among the topic proportions. Correlated topic model(CTM), where the topic proportions exhibit correlation via the logistic normal distribution, overcomes this limitation. Future work can include use of CTM based representation instead of LDA based representation for classification of images.